# A phenomenological multiscale framework for orientational interactions and viscoelasticity in migrating epithelial monolayers

Ivana Pajic-Lijakovic [1,2], Milan Milivojevic [1], and Peter V. E. McClintock[2]

[1] University of Belgrade, Faculty of Technology and Metallurgy, Department of Chemical Engineering, Belgrade, Serbia

[2] Department of Physics, Lancaster University, Lancaster LA1 4YB, UK

Correspondence to: Ivana Pajic-Lijakovic, iva@tmf.bg.ac.rs and

Peter McClintock, p.v.e.mcclintock@lancaster.ac.uk

**Abstract**

Collective migration of epithelial monolayers emerges from the interplay between mechanical interactions and biochemical signalling on the cellular scale. Here, we provide a phenomenological mechanobiological framework with constitutive interpretation linking microscopic orientational interactions to tissue-scale mechanics. By distinguishing reversible from irreversible head-on and glancing interactions, we find that reversible interactions store orientational mechanical energy while preserving the collision angle between cells before and after contact, whereas irreversible interactions dissipate energy and alter the collision angle. Quantifying the storage and dissipation of orientational energy is crucial, as it governs how collective migration and mechanical feedback emerge in epithelial monolayers, ultimately influencing density-dependent outcomes such as cell jamming and live cell extrusion. These interactions modulate cell elasticity, contractility, and adhesion, thereby shaping the surface properties and effective viscoelastic response of the monolayer and influencing the efficiency of collective cell migration. Increasing cell packing density shifts the balance from energy storage to dissipation, inducing density-dependent changes in migration mechanisms, epithelial surface tension, and viscoelasticity. We quantify these effects using orientational potentials, an effective second virial coefficient, and dimensionless ratios capturing the relative fractions of energy stored or dissipated by orientational interactions. While these contributions are partial at intermediate densities (above the confluent cell packing density and below the jamming packing density), they become dominant near jamming. This framework provides a mechanobiological perspective linking cell-scale interaction dynamics and orientation changes to emergent tissue-scale rheological behavior. It suggests how collision-induced orientational energy may contribute to collective migration mechanics and how density-dependent interaction regimes may influence the epithelial viscoelastic response under varying cell packing conditions.

**Glossary of terms**

**Cell jamming state transition**: the cell transition from the active (contractile) to the passive (non-contractile) state caused by an accumulation of cell compressive stress, leading to an increase in cell packing density. This process influences the viscoelasticity and surface characteristics of multicellular systems.

**Cell swirling motion:** a collective migration pattern in which cells move coherently along curved or circular trajectories, forming vortices or rotating streams within a tissue or cell monolayer.

**Cell unjamming state transition:** the transition from a mechanically arrested, densely packed state to a migratory state, driven by stress dissipation through weakening and remodeling of cell–cell and cell–matrix adhesions, which reduces compressive stress and cell packing density.

**Constitutive models**: stress-strain relationships.

**Epithelial cells**: exhibit cuboidal shape, limited mobility, apical-basal polarity, and strong E-cadherin-mediated cell-cell adhesions

**Epithelial surface tension**: a tension at the epithelial surface in contact with liquid along the tangential direction.

**Live cell extrusion**: occurs in areas where there are irregularities in the alignment of cells within the epithelial monolayers. These topological defects are caused by the combined effects of cell compressive and shear stress components.

**Mechanical stress**: a physical quantity that describes the magnitude and direction of the force per unit area that causes a deformation.

**Compressive stress**: a negative normal stress that acts inwards on the surface, tending to shorten or compact the material.

**Residual stress**: stress that remains in a system in the absence of external forces. It can be normal or shear. The residual stress can be either dissipative (viscous) or elastic.

**Shear strain**: deformation of a system in response to mechanical stress, applied tangentially.

**Shear stress**: stress that acts coplanar with the cross section of a system.

**Stress relaxation:** an observed, time-dependent, decrease in the stress of a system from an initial value towards the residual stress under constant strain, caused by structural ordering of the system under strain.

**Strain rate**: the change in strain per unit of time.

**Viscoelasticity**: a property of material, which defines the elastic and viscous characteristics of the system when it undergoes deformation.

## 1.Introduction

Epithelial monolayers are fundamental tissue architectures that coordinate collective cell migration during development, wound healing, and cancer invasion (Serra-Picamal et al., 2012; Shellard and Mayor, 2020). Cells within these monolayers exhibit mechanical coupling through adhesion junctions and cytoskeletal networks, allowing local forces to propagate across the tissue and influence large-scale behaviour (Rogers and Schier 2011; Collu and Mlodzik 2015; Aw and Devenport, 2017). Understanding how cells integrate mechanical interactions and biochemical signals is critical for explaining tissue morphogenesis and collective migration dynamics (Lang et al., 2024).

In migrating epithelial monolayers, mechanical interactions between cells play a central role in determining collective polarity and directional movement (Murray et al., 1988; Shellard and Mayor, 2020; Zegers et al., 2025). Rather than behaving as isolated units, epithelial cells remain mechanically integrated through E-cadherin-based adherens junctions, which not only maintain tissue cohesion but also act as conduits for transmitting mechanical signals and biasing cell polarity across the monolayer (Venhuizen and Zegers, 2017). Stable E-cadherin–mediated adhesion has been shown to be crucial for directional collective migration, with disruption of these junctions impairing coordinated re-orientation and migration efficiency. E-cadherin–mediated adhesion not only maintains tissue integrity but also transmits directional information and traction forces across the monolayer, contributing to collective orientation and migration directionality (Barriga and Mayor, 2019; Shellard and Mayor, 2020).

At the level of individual cell–cell contacts, distinct interaction geometries give rise to different orientational responses: head-on collisions tend to trigger contact inhibition of locomotion (CIL), leading to repolarization and weakening of cell–cell and cell–matrix adhesion (Roycraft and Mayor, 2016), whereas glancing interactions perturb alignment without complete repolarization, preserving junctional contacts and inducing localized torsional stress (Pajic-Lijakovic et al., 2024a). Moreover, compressive and shear stress components (**Glossary**) caused by collective cell migration can induce distinct orientational interactions — such as head-on and glancing collisions — that modulate cell realignment and polarity within the tissue (Saw et al., 2017; Alert and Trepat, 2020; Pajic-Lijakovic et al., 2024a). In contrast to cell head-on and glancing interactions, which have the potential to disrupt and even diminish directional cell migration, head-to-tail interactions among cells actually promote directional cell migration (Alert and Trepat, 2020). The mechanical response of the tissue depends on cell packing density, strength of cell-cell and cell-matrix adhesion contacts, and viscoelastic properties, giving rise to complex collective behaviour (Serra-Picamal et al., 2012; Pérez-González et al., 2019; Espina et al., 2023; Pajic-Lijakovic et al., 2024a). At high cell packing density, glancing interactions are statistically more likely than true head-on collisions, specially in the regions of higher cell shear stress. However, in the regions of epithelial monolayers where direct collisions between forward and backward flows take place, the head-on interactions primarily affect the rearrangement of cells (Pajic-Lijakovic et al., 2026).

In addition to mechanical interactions, cells respond to biochemical signals such as morphogens or polarity-regulating pathways (Aw and Devenport, 2017). Gradients of signaling activity provide spatial cues that bias cell orientation, allowing tissue-level coordination (Rogers and Schier 2011; Collu and Mlodzik 2015; Aw and Devenport, 2017). This signaling can propagate across the plane of the monolayer, forming an effective morphogen flux that couples biochemical information to mechanical orientation. The interplay between mechanical interactions and signaling gradients creates a mechano-chemical feedback loop that governs orientational energy storage and dissipation (Hannezo and Heisenberg, 2019).

A range of modelling frameworks has been developed to describe collective cell migration across scales. At the cellular level, Cellular Potts models provide a stochastic energy-minimization framework that captures cell–cell adhesion, area constraints, and rearrangements of cell configurations. These models have been widely applied to collective cell migration and can explicitly represent head-on cell–cell interactions, which may lead to contact inhibition of locomotion through local interaction rules (Graner and Glazier, 2016; Urcun et al., 2026; Alert and Trepat, 2020). Phase-field models represent cells as continuous fields governed by a free-energy functional and describe deformation and migration through evolving interfaces (Alert and Trepat, 2020; Yin et al., 2024). Vertex models describe epithelial tissues as networks of polygonal cells, focusing on junctional tension, cell shape, and topological rearrangements such as neighbour exchanges (Fletcher et al., 2014; Alt et al., 2017; Wang and Xu, 2023). These approaches provide detailed mechanistic descriptions but typically do not formulate tissue behaviour in terms of explicit macroscopic viscoelastic constitutive laws. At the tissue scale, poro-mechanical models describe coupled solid–fluid behaviour, accounting for interstitial fluid pressure, permeability, and transport, and have been increasingly used to describe and interpret experimental observations of tissue mechanics and fluid-mediated deformation (Lavigne et al., 2025a,b). In contrast, the present framework introduces a collision-level description of orientational interactions in collective migration, in which head-on and glancing encounters are classified either as reversible (effectively elastic) or as irreversible (dissipative) processes. The “orientational energy” in this work is an effective interaction potential defined at the coarse-grained cellular level, arising from mechanical contact interactions between cells, and should not be interpreted as biochemical or metabolic energy. This classification links interaction geometry to migration-induced stress generation and provides a bridge between encounter statistics and macroscopic viscoelastic response. At the continuum level, this is expressed as a regime-dependent constitutive description relating cell packing density, collective migration velocity, and stress generation.

In our framework, “reversible interactions” do not imply microscopic energy conservation. Cell reorientation involves active cytoskeletal remodelling and adhesion turnover, which are intrinsically dissipative processes. However, at the coarse-grained level of orientation dynamics, these processes are treated as effectively reversible if the system returns to the same mesoscopic state after interaction. Irreversibility corresponds to loss of recoverability in this reduced state space rather than absence of microscopic dissipation.

While morphogen gradients and mechanical coupling have been extensively studied separately, the quantitative integration of biochemical signalling, orientational energy, and collective mechanical response remains poorly understood. In particular, how head-on and glancing orientational interactions modulate both mechanical stress and signal propagation has not yet been fully captured within a unified framework. Here, we propose a theoretical framework linking orientational energy in migrating epithelial monolayers to local signalling activity. We introduce the concept of an informational flux as a phenomenological construct that couples mechanical orientation gradients to morphogen gradients, and we incorporate mechano-chemical feedback through orientation-dependent signalling activity. Within this framework, we explore how tissue-level orientational energy may be stored, dissipated, and modulated by biochemical signalling during collective migration. We analyse the coupled behaviour of this system across a range of packing densities and interaction geometries, highlighting how biochemical cues and mechanical interactions may jointly regulate collective cell orientation. This framework provides a mechanistic perspective on the integration of morphogen signalling and viscoelastic responses in epithelial tissues. (**Glossary**).

## 2. Biological aspects of cell-cell orientational interactions

In the context of migrating cells, cell–cell orientational interactions depend on the relative angle between the directions of migration of two neighboring cells. Head-on interactions, with collision angles near $\theta = \pi \pm \delta_\theta$, often lead to contact inhibition of locomotion (CIL) and repolarization. In glancing interactions at $\theta = \frac{\pi}{2} \pm \delta_\theta$, cells do not repolarize, but their trajectories and alignment can be perturbed by shear forces during brief contact (where $\delta_\theta$ is a small angle $\delta_\theta < \frac{\pi}{2}$) (Pajic-Lijakovic et al., 2024a). While head-on interactions are stimulated by an increase in compressive stress, glancing interactions are induced by the interplay between compressive and shear stress components. Compressive stress leads to an increase in cell packing density, while shear stress has no impact on the packing density, but has the potential to generate topological defects of cell alignment and can also induce cell swirling motion (**Glossary**) (Saw et al., 2017; Lin et al., 2018; Pajic-Lijakovic et al., 2024a). While Madin Darby Canine Kidney (MDCK) cells create full swirls, epithelial HaCaT skin cells and Caco2 intestinal cells exhibit movement in a uniform direction without central symmetry. The only aspect that changes over time is the collective direction of motion, which results in circular paths resembling incomplete swirls, as documented by Peyret et al. (2019). It is consistent with the observation that MDCK cells exhibit weaker cell-cell adhesions when compared to skin epithelial cells in a confluent state (Petrolli et al., 2021).

Head-on interactions occur when two migrating cells approach along roughly opposing directions, leading to collisions between their leading edges. These collisions generate significant compressive force at the point of contact. Such events often engage strong cytoskeletal interaction and adhesion dynamics. The canonical response to head-on collisions is contact inhibition of locomotion — cells halt forward movement at the site of contact, retract lamellipodia or filopodia at the collision interface, and repolarise their migration machinery to move away from each other (Mayor and Carmona-Fontaine, 2010; Zimmermann et al., 2016). This involves rearrangement of the actin cytoskeleton and cell polarity machinery — such as inhibition of Rac1 at the contact site and activation of RhoA, which drives contraction and retraction (Bischoff et al., 2021). Contact at the leading edge transiently reduces the strength of adherens junctions AJs and focal adhesions FAs (E-cadherin-mediated cell-cell adhesions and integrin-mediated cell-matrix adhesions) during the repolarization and disengagement process (Mayor and Roycraft, 2016). Both types of adhesion complexes, AJs and FAs, are intracellularly connected to the actin cytoskeleton and activate a unified set of signalling proteins and actin regulators, including Rho family GTPases (Mui et al., 2016). At the contact site, E-cadherin-mediated junctions transiently weaken, allowing the leading edge to retract and the cell to move away. RhoA is activated at the contact site (contractility), Rac1 is inhibited (protrusion suppression) (Mayor and Carmona-Fontaine, 2010). Compressive stress is sensed through E-cadherin junctions, which link to the actin cytoskeleton and can influence YAP/TAZ nuclear translocation (Kim et al., 2011).

Glancing interactions preserve strong E-cadherin-mediated cell-cell contacts but generate torsional shear stress at the cell-matrix biointerface. This stress leads to conformational changes in integrin molecules, which can cause the detachment of FAs and may trigger a signaling cascade such as S1P signaling facilitated by ROCK-mediated actomyosin contraction, which is necessary for the extrusion of live cells (Eisenhoffer et al., 2012). The detachment of an FA can be caused by interfacial shear stress between the cell and the matrix, usually ranging from 4 to 6 Pa (Paddillaya et al., 2019). In turn, torsional stress can activate ERK and calcium signaling pathways, which encode the spatial and temporal information needed for gap repair or collective cell response. These interactions generate asymmetric forces along cell edges without inducing full repolarization. Unlike head-on collisions that typically induce contact inhibition of locomotion and a polarity switch, glancing interactions do not induce repolarization of the migrating cell. Cells largely keep their existing front–rear polarity after these encounters and strong cell-cell adhesion contacts (Lin et al., 2018; Pajic-Lijakovic et al., 2024a). Although glancing interactions do not cause CIL, they can have other tissue-level consequences such as: (i) inhomogeneous alignment and stress distribution and (ii) live cell extrusion (**Glossary**) in crowded regions. Perturbation of cellular alignment causes an imbalance in the intercellular forces that are crucial for cell realignment, ultimately leading to the restoration of force equilibrium (Chen et al., 2018).

Cells can actively self-regulate the strength of adhesion complexes through the turnover of cadherin and integrin molecules, which is influenced by the level of mechanical stress that accumulates within the cells (Iyer et al., 2019). Cadherin and integrin exhibit a turnover time of just a few minutes (Lee and Wolgemuth, 2011). As cells move, integrin clusters attach to the substrate matrix, mature, and then detach in succession. The lifespan of FAs is typically several tens of minutes (Stehbens and Wittmann, 2014). This dynamic process of FA attachment and detachment during cellular movement helps to protect the cell from the accumulation of shear stress between the cell and the matrix.

These head-on and glancing interactions transiently store orientational energy and subtly influence the cell's mechanical state through remodeling of the actin cytoskeleton and adhesion complexes, which act as "mechanical switches" that convert physical cues into biological responses. Together with head-on interactions, glancing interactions couple orientational energy storage and dissipation to dynamic remodelling of cell elasticity, contractility, and cohesion, allowing isotropic migration and density-dependent modulation of viscoelasticity and epithelial surface tension (Pajic-Lijakovic et al., 2023; 2026). Depending on the extent to which interactions related to cell orientation lead to dissipation of orientational energy, we can categorize cell-cell interactions based on orientation reversibility as either reversible or irreversible.

## 3. Cell-cell reversible and irreversible orientational interactions

Cell-cell orientational interactions dominantly influence the viscoelasticity caused by collective cell migration under higher surface cell packing densities $n_e$, i.e. $n_e > n_{conf}$(where $n_{conf}$ is the cell packing density in the confluent state). Petitjean et al. (2010) revealed that the MDCK cell monolayer reached the confluence for a cell packing density of $n_{conf} \sim 2.5x10^5 \ \frac{cells}{cm^2}$ and velocity of $\sim 0.14 \ \frac{\mu m}{min}$.

First, it is necessary to define the orientational angle between cells before and after collision. The criterion for reversible orientational interactions between cells is defined by the preservation of the relative polarity angle, i.e., $\theta_{12}^- = \theta_{12}^+$ (where $\theta_{12}^-$ and $\theta_{12}^+$ are the angles between cells 1 and 2 before and after collision such that the angle $\theta_{12}^-$ is equal to $\theta_{12}^- = \theta_1^- - \theta_2^-$ and the angle $\theta_{12}^+$ is equal to $\theta_{12}^+ = \theta_1^+ - \theta_2^+$, $\theta_1^-$ and $\theta_1^+$ are the angles of cell 1 before and after collisions relative to the direction of migration, while $\theta_2^-$ and $\theta_2^+$ are the angles of cell 2 before and after collisions relative to the direction of migration). The angle $\theta_i^-$ and $\theta_i^+$correlates with cell velocities $\vec{\boldsymbol{v}}_{\boldsymbol{i}}^-$ and $\vec{\boldsymbol{v}}_{\boldsymbol{i}}^+$, respectively $i = 1{,}2$ and the mesoscopic velocity within a multicellular domain $\vec{\boldsymbol{v}}(r,\tau)$. The coarse-grained cell velocity at $r$ can be expressed as: $\vec{\boldsymbol{v}}(r,\tau) = \frac{\Delta A_m}{N_c} \sum_i^{N_c} \vec{\boldsymbol{v}}_{\boldsymbol{i}} \delta(r - r_i)$ (where $N_c$ is the number of cells located at $r$ at time $\tau$, $\Delta A_m$ is the increment of the surface area of epithelial monolayer, $\vec{\boldsymbol{v}}_{\boldsymbol{i}}$ is the velocity of the *i*-th cell located at $r_i$, and $\delta(\cdot)$ is a delta function). In practice, this field is evaluated using a finite coarse-graining procedure in which the Dirac delta is replaced by a normalized kernel with characteristic width comparable to the cell diameter (**Appendix A**). The velocity field is evaluated over a finite temporal coarse-graining window $\Delta\tau$, chosen to be comparable to the characteristic persistence time of cell motion in order to suppress short-time fluctuations while retaining collective migration dynamics. Consequently, angles satisfy the condition that:

$$\cos\theta_i^- = \frac{(\vec{\boldsymbol{v}}_i^- - \vec{\boldsymbol{v}}) \cdot \vec{\boldsymbol{n}}_i^- \cdot}{|\vec{\boldsymbol{v}}_i^- - \vec{\boldsymbol{v}}|} \text{ and } \cos\theta_i^+ = \frac{(\vec{\boldsymbol{v}}_i^+ - \vec{\boldsymbol{v}}) \cdot \vec{\boldsymbol{n}}_i^+}{|\vec{\boldsymbol{v}}_i^+ - \vec{\boldsymbol{v}}|} \quad (1)$$

where $\vec{\boldsymbol{n}}_{\boldsymbol{i}}^-$ and $\vec{\boldsymbol{n}}_{\boldsymbol{i}}^+$ are unit vectors equal to $\vec{\boldsymbol{n}}_{\boldsymbol{i}}^- = \frac{\vec{\boldsymbol{r}}_i^- - \vec{\boldsymbol{r}}}{|\vec{\boldsymbol{r}}_i^- - \vec{\boldsymbol{r}}|}$ and $\vec{\boldsymbol{n}}_{\boldsymbol{i}}^+ = \frac{\vec{\boldsymbol{r}}_i^+ - \vec{\boldsymbol{r}}}{|\vec{\boldsymbol{r}}_i^+ - \vec{\boldsymbol{r}}|}$ . A normalized projection of the velocity vector difference onto the unit vectors is used to avoid the angular branch cut. An indicator that orientational interactions are irreversible is: $\theta_{12}^- \neq \theta_{12}^+$. In this case, part of the orientational energy is dissipated. The collision angle $\theta \equiv \theta_{12}^+ = \theta_1^+ - \theta_2^+$ is equal to: (i) $\theta = \pi \pm \delta_\theta$ corresponding to cell head-on interactions and (ii) $\theta = \frac{\pi}{2} \pm \delta_\theta$ corresponding to glancing interactions and $\delta_\theta$ is the small angle $\delta_\theta < \frac{\pi}{2}$. To facilitate numerical computation, the angular tolerance is fixed at $\delta_\theta = \frac{\pi}{6}$. Irreversible interactions are characterized by a change in relative orientation, $\theta_{12}^- \neq \theta_{12}^+$ (where the post-collision angle $\theta_{12}^+$ is constrained within the specified head-on or glancing bins. This shift represents a loss of orientational order and a dissipation of energy during the remodeling process. While head-on interactions cause cell repolarisation accompanied by a weakening of the cell-cell and cell-matrix adhesion contacts, glancing interactions induce mechanical rotation along the axis of cell migration and generate torsion stress along focal adhesion (cell-matrix adhesion contact) (Roycraft and Mayor, 2016; Pajic-Lijakovic et al., 2024a).

During both head-on and glancing interactions, cells undergo rotation and realignment, which in reality involves frictional effects and transient energy dissipation within the cytoskeleton and adhesion machinery. However, in the present framework, we focus on the cell's ability to restore its original relative polarity after the interaction, such that the net change in orientation vanishes. This definition captures the reversible storage of orientational energy without introducing net dissipation, while acknowledging that transient rotational friction is an inherent physical process during reorientation. In all cases, translational kinetic energy is negligible. Consequently, cell-cell interactions can have eight reversible/irreversible outcomes as shown in **Table 1**:

**Table 1**. Outcomes of cell-cell orientational interactions

| type | Cell 1 | Cell 2 | Physical meaning |
|---|---|---|---|
| reversible head-on interactions (E) | move $\theta_1^+ = \theta_1^- + \pi$ | move $\theta_2^+ = \theta_2^- + \pi$ | pure reorientation, no dissipation $\theta_{12}^- = \theta_{12}^+$ Recoverable reorientation in orientational state space; collision angle preserved. Microscopic dissipation may still occur due to active remodeling. |
| irreversible head-on interactions (NE-11) | move $\theta_1^+ = \theta_1^- + \pi \pm \delta_1$ | move $\theta_2^+ = \theta_2^- + \pi \pm \delta_2$ $\delta_1 \neq \delta_2$ | remodeling but no arrest; perturbation of cell alignment $\theta_{12}^- \neq \theta_{12}^+$ Partial loss of orientational correlation; asymmetric remodeling of alignment. Microscopic dissipation accompanies reorientation. |
| irreversible head-on interactions (NE-10) | move $\theta_1^+ = \theta_1^- + \pi \pm \delta_1$ | stop | asymmetric arrest; one cell stops, strong dissipation Asymmetric arrest leading to loss of orientational memory and increased dissipation at the coarse-grained level. |
| irreversible head-on interactions (NE-01) | stop | move $\theta_2^+ = \theta_2^- + \pi \pm \delta_2$ | asymmetric arrest; one cell stops, strong dissipation Asymmetric arrest leading to loss of orientational memory and increased dissipation at the coarse-grained level. |
| irreversible head-on interactions (NE-00) | stop | stop | Mutual arrest; strong loss of orientational order; precursor to jamming. |
| reversible glancing interactions (E) | $\theta_1^+ = \theta_1^- + \frac{\pi}{2} - \delta$ | $\theta_2^+ = \theta_2^- + \frac{\pi}{2} - \delta$ | no dissipation $\theta_{12}^- = \theta_{12}^+$ Recoverable perturbation of orientation; alignment restored in reduced state description; microscopic dissipation still present. |
| irreversible glancing interactions (NE-11) | $\theta_1^+ = \theta_1^- + \frac{\pi}{2} - \delta_1$ | $\theta_2^+ = \theta_2^- + \frac{\pi}{2} - \delta_2$ $\delta_1 \neq \delta_2$ | perturbation of cell alignment $\theta_{12}^- \neq \theta_{12}^+$ Loss of orientational correlation due to asymmetric remodeling. |
| irreversible (NE-X) | extruded | stop/migrate | live cell extrusion triggered by high density; topology change Topological change with loss of orientational description validity; strongly dissipative at cellular level. |

**Note:** All angular additions and subtractions are performed $mod 2\pi$ to ensure continuity across the angular branch cut. This ensures that $\theta_{12}^-$, $\theta_{12}^+ \in [0,2\pi)$. The stochastic variables $\delta_1$ and $\delta_2$ represent the angular noise or remodelling magnitude for each cell. To reflect the loss of orientational correlation in irreversible interactions, these values are sampled such that $\delta_1 \neq \delta_2$, ensuring a non-zero change in the relative orientation. For consistency with the defined bins, these satisfy $|\delta_i| < \delta_\theta$, where $i \equiv 1,2$ and $\delta_\theta = \frac{\pi}{6}$. Reversibility in this table refers to recoverability of the coarse-grained orientational state (collision angle and alignment) and does not imply thermodynamic or energetic reversibility. All cellular processes involve active remodeling and energy consumption; therefore, microscopic dissipation is present in all interaction types.

While the current framework provides a computable basis for cell-cell interactions, future implementations could benefit from advanced numerical refinements. For instance, posteriori error estimation and dual-weighted residual refinement (Bui et al., 2025) offers robust pathways for quantity-of-interest validation. Furthermore, the integration of physics-informed finite-element approaches (Xiong et al., 2025) or mixed-dimensional coupling strategies (Sky et al., 2024) could enhance the modeling of multi-scale interactions between discrete cell dynamics and continuous tissue-level dynamics.

For a given angle of collision $\theta$ and cell packing density $n_e$, several normalized probabilities can be obtained for both types of interactions: $P_E(\theta, n_e)$, $P_{NE-11}(\theta, n_e)$, $P_{NE-10}(\theta, n_e)$, $P_{NE-01}(\theta, n_e)$, $P_{NE-00}(\theta, n_e)$, and $P_{NE-X}(\theta, n_e)$ that satisfy the condition that: $P_E + P_{NE-11} + P_{NE-10} + P_{NE-01} + P_{NE-00} + P_{NE-X} = 1$. Reversible and irreversible head-on interactions, represented as: NE-10, 01, and 00, are shown schematically in **Figure 1**:

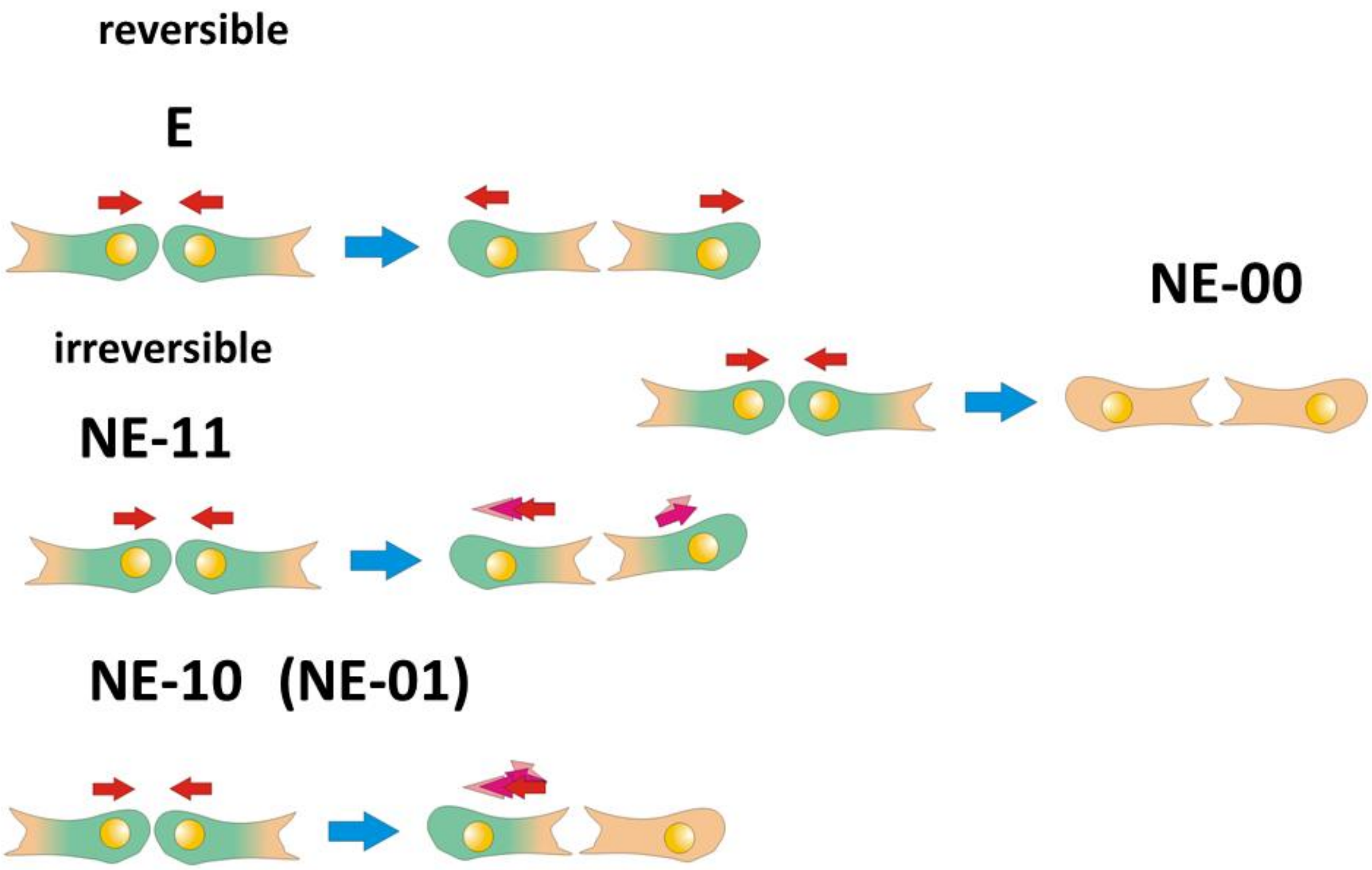

**Figure 1**. Schematic presentation of reversible (E) and irreversible (NE) head-on interactions, represented as: NE-11, 10, 01, and 00 as defined in **Table 1**. Red arrows indicate cell orientation before and after cell-cell collision, while pink arrows indicate a set of possible directions. Blue arrows connect cell states before and after collision. Partially green labeled cells are migrating cells, while brown cells are non-migrating cells. These non-migratory cells can start migration again during cell unjamming (**Glossary**). Green parts of migrating cells represent their front sides. Head-on interactions trigger cell repolarization. When cells have insufficient time to repolarize between two head-on interactions, one or both of them can undergo jamming (i.e., the contractile-to-non contractile cell state transition).

These irreversible outcomes (NE-10, 01, and 00) are mainly linked to head-on cell interactions instead of glancing interactions, primarily due to the fact that direct interactions trigger cell repolarization (Roycraft and Mayor, 2016). When the interval between two head-on collisions is shorter than the duration of cell repolarization, cells have insufficient time to finish the repolarization process and experience a transition from the contractile, migrating state to a non-contractile, resting state leading to a few outcomes NE-10, NE-01, or NE-00 (Pajic-Lijakovic et al., 2024a). The phenomenon of cell contractile-to-non contractile transition in an overcrowded environment is known as cell jamming (**Glossary**). The average repolarization time during the rearrangement of confluent epithelial Madin Darby Canine Kidney (MDCK) cell monolayers is $1.28\ \mathrm{h}$ (Norbohm et al., 2016). However, this time can be extended in an overcrowded environment (Pajic-Lijakovic et al., 2024a).

Reversible and irreversible NE-11 and NE-X glancing interactions are shown schematically in **Figure 2**:

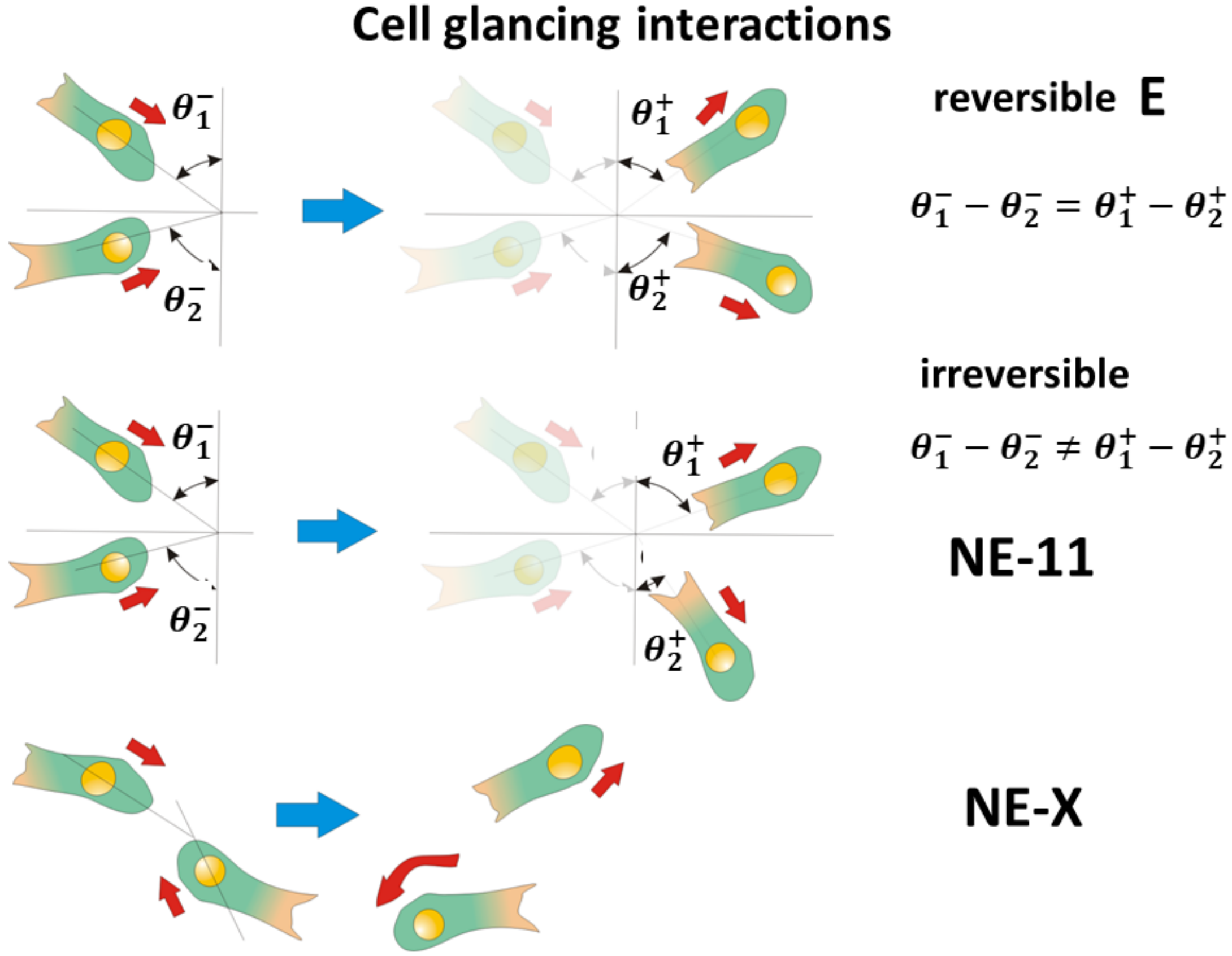

**Figure 2**. Schematic presentation of reversible (E) and irreversible NE-11 and NE-X glancing interactions. Red arrows indicate the direction of cell migration, i.e., cell orientations before and after the cell-cell collision. Blue arrows connect the cell state before and after the collision. Green parts of migrating cells represent the fronts of cells. Cells retain their active contractile state after glancing interactions. Torsional stress caused by glancing interactions can induce the disruption of FAs leading to live cell extrusion (NE-X).

The glancing, irreversible interaction NE-X occurs primarily in high-density regimes under shear and compressive stresses where lateral sliding is blocked and the torsional shear stress generated disrupts focal FAs (Saw et al., 2017; Pajic-Lijakovic et al., 2024a). Consequently, for head-on and glancing interactions, the probabilities satisfy the conditions that: $P_E^h + P_E^g + P_{NE-11}^h + P_{NE-11}^g + P_{NE-10} + P_{NE-01} + P_{NE-00} + P_{NE-X} = 1$. The outcomes that are included, as well as their extent, are contingent upon the angle of collision and the density of cell packing, which are recognized as the main control factors (Pajic-Lijakovic et al., 2024a).

3.1 **The characterisation of cell packing density regimes: cell-cell interactions**

The main characteristic of migrating epithelial monolayers, which has been experimentally confirmed, is inhomogeneous distributions of: cell packing density, cell velocity, corresponding strain, and cell mechanical stress (Serra-Picamal et al, 2012; Nnetu et al., 2013; Notbohm et al., 2016;Tlili et al., 2018; Pérez-González et al., 2019; Pajic-Lijakovic et al., 2026). Consequently, a cell monolayer can be treated as an ensemble of multicellular mesoscopic domains. Domains are unstable. They exist for some period of time and then lose their identity due to interactions between domains as shown in **Figure 3**.

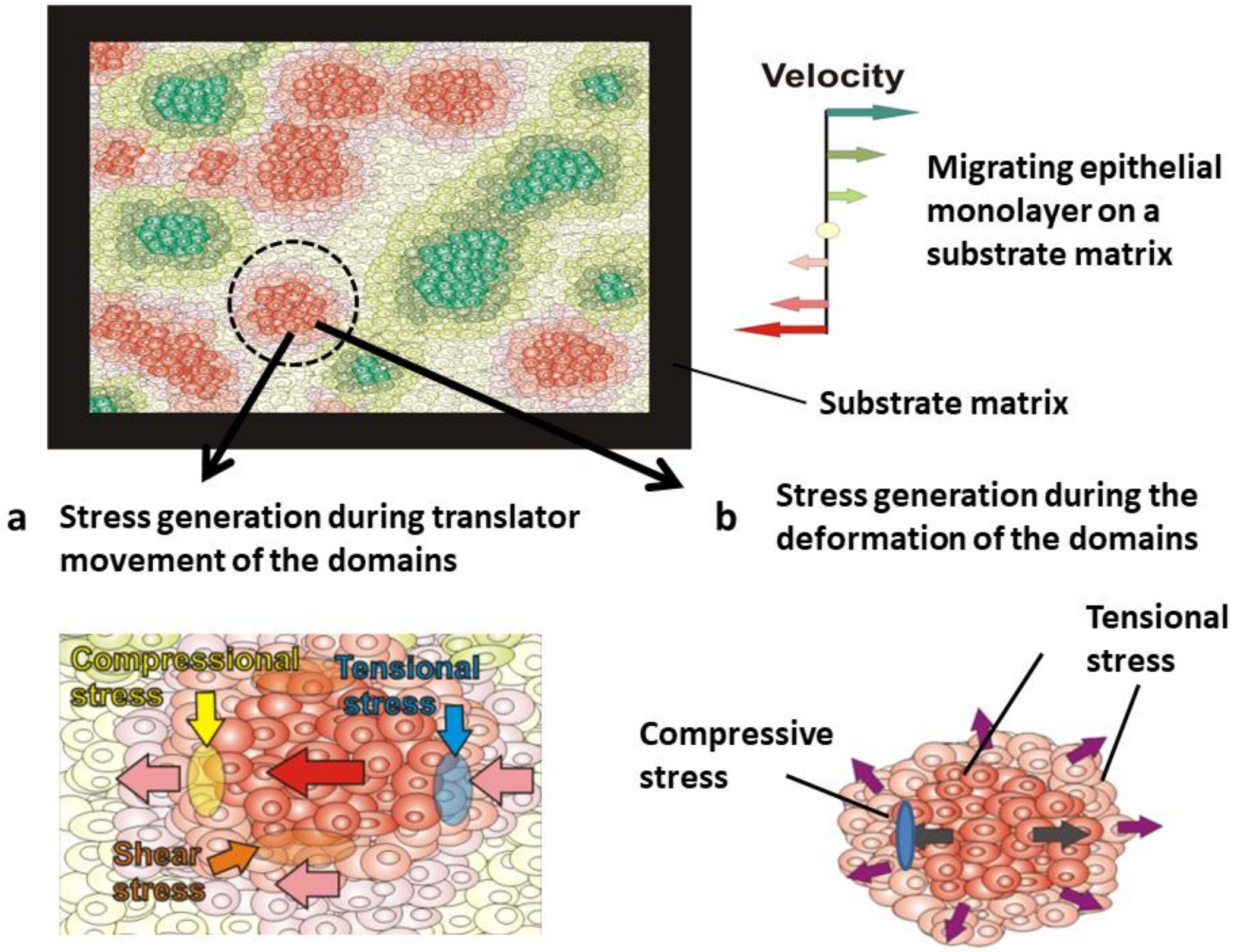

**Figure 3**. Collective migration of epithelial monolayer on a substrate matrix, which is inspired by experimental data from Serra-Picamal et al. (2012). Domains undergo extension/compression and translation movements, which generate mechanical stress. (a) Translation of domains at higher speeds than those of surrounding domains generates all stress components. Shear stress can be intensive along the biointerface. (b) Extension/compression of the domain generates tensional/compressive stress within it and generation of compressive/tensional stress along the biointerface with the surrounding domain, which is extended/compressed slowly. (Yellow, blue, and orange arrows indicated the places where stress is generated in **Figure 3a**, while red and pink arrows show the direction of domain translation. Red domains migrate faster than the surrounding pink domains. Black and purple arrows indicate extensions of the red and pink domains presented in **Figure 3b**. The extension of red domain is more intensive than that of pink domain. Consequently, the red domain pushes on the pink domain leading to the generation of compressive stress along the biointerface between the domains.

During its existence, a domain can be characterised by homogeneous distributions of relevant physical parameters. Several density-dependent domain regimes, which influence epithelial surface tension and viscoelasticity, can be established. Domains undergo extension/compression and translation. The extension/compression of some domains is more intensive than for others (**Figure 3a**). It depends on the relationship between the adhesion and cohesion energies per unit surface, which vary along the monolayer (Serra-Picamal et al., 2012; Pérez-González et al., 2019). Consequently, normal stress (compressive and tensional) is generated within the domains and during the interactions between domains themselves, while shear stress is generated primarily along the biointerface between the domains (Serra-Picamal et al., 2012; Saw et al., 2017; Pajic-Lijakovic et al., 2026). Translation of the domains at various speeds generates mechanical stress as shown in Figure 3a. Collision between the domains that undergo forward and backward movement generates compressive stress. Cells tend to align with the tensile direction, following the path of maximum principal stress (Trepat and Fredberg, 2011). The phenomenon is known as plithotaxis. However, orientatiomal interactions perturb cell alignment.

An increase in cell-packing density, which occurs over hours, stimulates cell-cell interactions. Consequently, three cell-packing density regimes can be distinguished based on experimental data from the literature (Pajic-Lijakovic et al., 2025; 2026), which is shown schematically in **Figure 4**:

**Cell packing density regimes**

| cell packing density | velocity | characteristics |
|---|---|---|
| $n_e \leq n_{conf}$ | $0.1 < \vec{v}_c < \sim 1 \frac{\mu m}{min}$ | **anisotropic migration** |
| $n_{conf} < n_e < n_j$ | $\vec{v}_c \sim 10^{-3} - 10^{-2} \frac{\mu m}{min}$ | **isotropic migration** |
| $n_e \to n_j$ | $\vec{v}_c \to 0$ | **damped migration near cell jamming** |

Velocity

$v_x$

**Figure 4**. Schematic representation of the key characteristics of distinct cell-packing density regimes: low, intermediate, and high. Cell-packing density modulates the nature of cell–cell orientational interactions, which in turn feedback on cell migration mechanisms, leading to a transition from convective to diffusive and ultimately to sub-diffusive behaviour.

- **Lower cell-packing density regime:** for $n_e \leq n_{conf}$, where $n_{conf} \sim 2.5x10^5 \frac{\text{cells}}{\text{cm}^2}$ is the surface cell-packing density in the confluent state, while the corresponding cell velocity is $0.1 < \vec{\boldsymbol{v}}_{\boldsymbol{c}} < \sim 1 \frac{\mu\text{m}}{\text{min}}$ (Pajic-Lijakovic and Milivojevic, 2025)
  Cell migration is anisotropic and corresponds to a convective mechanism (Pajic-Lijakovic et al., 2026). Cell-cell positional interactions, rather than cell-cell orientational interactions, dominate cell rearrangement. The frequency of cell-cell orientational interactions tends to zero, i.e., $f_c \rightarrow 0$. The velocity correlation length is an order of magnitude higher than the average size of a single cell (Petrolli et al., 2021; Wang and Xu, 2023).
- **Intermediate cell-packing density regime:** for $n_{conf} < n_e < n_j$, where $n_j$ is the cell packing density in the jamming state, which is an order of magnitude higher than $n_{conf}$ (**Glossary**), while the corresponding cell velocity is $\vec{\boldsymbol{v}}_{\boldsymbol{c}} \sim 10^{-3} - 10^{-2} \frac{\mu\text{m}}{\text{min}}$ (Nnety et al., 2013)
  Here, cell-cell orientational interactions dominate cell rearrangement. These interactions have the potential to perturb aligned cell migration. The velocity correlation length decreases in this regime (Petrolli et al., 2021; Wang and Xu, 2023). Cell migration is isotropic and corresponds to a diffusion mechanism (Pajic-Lijakovic and Milivojevic, 2025). The frequency of cell-cell orientational interactions $f_c$ and fraction of irreversible head-on and glancing interactions NE-11 increases with cell packing density, while the other irreversible interactions such as: NE-10, NE-01, NE-00, and NE-X can be neglected. It means that cells resume migration after reversible/irreversible interactions. The frequency $f_c < \frac{1}{\tau_{RP}}$ (where $\tau_{RP}$ is the average cell repolarisation time). Consequently, cells have enough time to repolarise after collisions.
- **High cell-packing density regime near jamming:** for $n_e \rightarrow n_j$, while the corresponding cell velocity $\vec{\boldsymbol{v}}_{\boldsymbol{c}} \rightarrow 0$ (Nnetu et al., 2013; Pajic-Lijakovic and Milivojevic, 2025).
  Cell-cell orientational interactions dominate cell rearrangement. Cell migration is damped, isotropic, and corresponds to a sub-diffusion mechanism. The velocity correlation length is equal to the size of a single cell (Petrolli et al., 2021). The frequency of cell-cell orientational interactions is higher than the threshold, i.e., $f_c \geq \frac{1}{\tau_{RP}}$ and the fraction of irreversible interactions NE-10, NE-01, NE-00, and NE-X tends to unity. In this regime, the majority of irreversible cell-cell interactions cause one or both cells stop their migration or eliminate the cell from the monolayer (live cell extrusion). It is in accordance with the fact that cells have insufficient time to repolarise between two collisions.

Changes in cell orientational energy, arising from head-on and glancing interactions, have a profound impact on the mechanical and migratory state of epithelial cells. When cells undergo NE 11 interactions, elastic orientational energy is stored transiently in their relative orientation, while dissipation occurs through polarity reorientation and the remodelling of cortical tension, contractility, and cell-cell and cell-matrix adhesion complexes. This orientational energy influences cell velocity, migration direction, and the local distribution of cytoskeletal tension, effectively modulating the cell's elastic and contractile state, as well as its cohesive interactions with neighbours (Rausch et al., 2013; Chen et al., 2018). In the intermediate-density regime, these orientational interactions contribute partially to the total viscoelastic energy of the monolayer, whereas in high-density regimes, repeated collisions dominate both energy storage and dissipation, strongly constraining cell motion (Pajic-Lijakovic et al., 2025). Consequently, orientational energy acts as a key regulator linking intercellular interactions to the dynamic modulation of cell mechanics and tissue-level viscoelasticity.

The proposed density-dependent regimes can be characterised based on the dimensionless parameter:

$$\lambda = \tau_{RP} f_c \tag{2}$$

where $\tau_{RP}$ is the average repolarisation time and $f_c$ is the collision frequency. The collision-repolarisation number $\lambda$ takes values: (i) $\lambda \rightarrow 0$ for the low-density regime, (ii) $\lambda \rightarrow 1$ for the intermediate-density regime, and (iii) $\lambda \gg 1$ for the high-density regime. In further consideration, it is necessary to derive expressions for the storage and dissipation of orientational energy.

### 3.2 Energy storage and energy dissipation during cell-cell orientational reversible and irreversible interactions

In dense epithelial monolayers head-on and glancing interactions induce storage and dissipation of cell orientational potential energy. These have a feedback impact on overall storage and dissipation of elastic, contractile and cohesive energy within migrating epithelial monolayers. Head-on interactions: (i) trigger cell repolarisation, which influences the cells' elasticity and contractility and (ii) weaken cell-cell and cell matrix adhesion contacts leading to energy dissipation (Roycraft and Mayor, 2016). Glancing interactions are related to cells sliding over each other, which causes: (i) elastic deformation of cortex and cell-cell and cell-matrix adhesion contacts and (ii) energy dissipation caused by frictional or viscoelastic relaxation along the sliding contact (Farrell et al., 2012; Smeets et al., 2016).

Reversible and irreversible cell–cell interactions can be quantified by the effective angular moduli of the collisions $k_E$ and $k_{NE}$ respectively (Farell et al., 2012; Smeets et al., 2016), which account for the strength of adhesion contacts and cortical stiffness. Two types of angular moduli can be distinguished: (i) normal moduli for head-on interactions $k_E^n$, $k_{NE-11}^n$, $k_{NE-10}^n$, $k_{NE-01}^n$, and $k_{NE-00}^n$, and (ii) shear moduli for glancing interactions $k_E^s$, $k_{NE-11}^s$, and $k_{NE-X}^s$. It aligns with the understanding that head-on interactions between cells are primarily triggered by an increase in compressive stress, which results in a higher packing density of cells (Pajic-Lijakovic et al., 2024a). Conversely, glancing interactions are influenced by shear stress, although an increase in cell packing density also plays a significant role in the overall effect (Saw et al., 2017; Pajic-Lijakovic et al., 2024a). Reversible collisions store energy fully and reversibly, while irreversible collisions store energy partially and dissipate the rest, with the magnitude depending on the collision outcome and repolarisation dynamics. The free energy of cell repolarisation is equal to $\sim 10^{-12}$ J (Zhong et al., 2014), which is the same order of magnitude as the work of the cell traction force per single cell (Butler et al., 2002). Consequently, the moduli of head-on interactions satisfy the conditions that: (i) $k_E^n > k_{NE-11}^n > k_{NE-10}^n$, (ii) $k_{NE-10}^n = k_{NE-01}^n$, and (iii) $k_{NE-00}^n, = 0$, while the moduli of glancing interactions satisfy the conditions that: (i) $k_E^s > k_{NE-11}^s$ and (ii) $k_{NE-X}^s = 0$.

### 3.2.1 Energy storage and dissipation in single orientational interactions

Farrell et al. (2012) proposed a harmonic potential for cell-cell orientational interactions. In this line, the storage energy per single reversible head-on interaction is equal to: $U_E^{stor-h}(\theta) = \frac{1}{2} k_E^n (\delta\theta)^2$, while the storage energy of irreversible interaction is $U_{NE-i}^{stor-h}(\theta) = \frac{1}{2} k_{NE-i}^n (\delta\theta)^2$ (where $i \equiv 11,10,01,00$). The energy dissipation per single head-on interaction is $U_{NE-i}^{diss-h}(\theta) = U_E^{stor-h}(\theta) - U_{NE-i}^{stor-h}(\theta)$, such that $U_{NE-i}^{diss-h}(\theta) > 0$.

The storage energy per single elastic glancing interaction is: $U_E^{stor-g}(\theta) = \frac{1}{2} k_E^s (\delta\theta)^2$, while the storage energy of irreversible interaction is $U_{NE-i}^{stor-g}(\theta) = \frac{1}{2} k_{NE-i}^s (\delta\theta)^2$ (where $i \equiv 11, X$). The energy dissipation per single glancing interaction is $U_{NE-i}^{diss-g}(\theta) = U_E^{stor-g}(\theta) - U_{NE-i}^{stor-g}(\theta)$, such that $U_{NE-i}^{diss-g}(\theta) > 0$. The cumulative effects of cell-cell interactions lead to energy storage and dissipation within the monolayers.

### 3.2.2 Energy storage and dissipation during orientational interactions within epithelial monolayers

The storage surface energy density caused by head-on interactions during a time increment $\Delta\tau$ can be expressed as: $\Delta E_{stor}^h = \frac{1}{2} f_c \Delta\tau \int_{\Omega_h} \left[ P_E^h U_E^{stor-h}(\theta) + \sum_i P_{NE-i}^h U_{NE-i}^{stor-h}(\theta) \right] \rho(\theta,\tau) d\theta$, while the storage surface energy density caused by glancing interactions is: $\Delta E_{stor}^g = \frac{1}{2} f_c \Delta\tau \int_{\Omega_g} \left[ P_E^h U_E^{stor-g}(\theta) + \sum_i P_{NE-i}^g U_{NE-i}^{stor-g}(\theta) \right] \rho(\theta,\tau) d\theta$ (where $\Omega_h \equiv [\pi - \delta_h, \pi + \delta_h]$, $\Omega_g \equiv \left[\frac{\pi}{2} - \delta_g, \frac{\pi}{2} + \delta_g\right]$, $\delta_h$ and $\delta_g$ are small angles, while $\rho(\theta)$ is the collision angle distribution satisfying the condition that: $\int_0^\pi \rho(\theta,\tau) d\theta = 1$) (Farell et al., 2012). The distribution $\rho(\theta)$ for isotropic cell migration can be simplified as: $\rho(\theta) \sim \frac{1}{\pi}$. Energy dissipation caused by irreversible head-on interactions per unit surface in the time increment $\Delta\tau$ is equal to: $\Delta E_{diss}^h = \frac{1}{2} f_c \Delta\tau \int_{\Omega_h} [(P_E^h + \sum_i P_{NE-i}^h) U_E^{stor-h}(\theta) - \sum_i P_{NE-i}^h U_{NE-i}^{stor-h}(\theta)] \rho(\theta,\tau) d\theta$, while the energy dissipation caused by glancing interactions is: $\Delta E_{diss}^g = \frac{1}{2} f_c \Delta\tau \int_{\Omega_g} [(P_E^g + \sum_i P_{NE-i}^g) U_E^{stor-g}(\theta) - \sum_i U_{NE-i}^{stor-g}(\theta)] \rho(\theta,\tau) d\theta$.

### 3.2.3 The effective angular modulus for head-on and glancing interactions

The effective angular modulus for cell head-on interactions expressed as: $k_{eff}^h(\theta, n_e) = P_E^h k_E^n + \sum_i P_{NE-i}^h \, k_{NE-i}^n$ and the effective angular modulus for cell glancing interactions expressed as: $k_{eff}^g(\theta, n_e) = P_E^g k_E^s + \sum_i P_{NE-i}^g \, k_{NE-i}^s$ can be related to the 2D second virial coefficient. The storage and dissipation of orientational energy activate a cell signaling cascade which has a feedback impact on the cell-cell interactions themselves.

### 3.3 Control of orientational energy storage/dissipation by morphogen flux

The storage/dissipation of orientational energy during head-on and glancing interactions is controlled by the morphogen flux. Morphogens are biochemical signals that provide spatial information to cells, influencing their polarity, orientation, or fate (Rogers and Schier, 2011). Here, we consider morphogens as coarse-grained, dimensionless signalling activity $a(r,\tau) \in [0,1]$ representing the local activation level of polarity-regulating pathways (e.g., PCP, Rac/Rho). Gradients of morphogen activity can bias cell polarity and planar orientation within epithelial monolayers, with core planar cell polarity machinery interpreting these gradients into cellular asymmetry. In collective migration, signalling pathways such as planar cell polarity intersect with mechanical forces to coordinate cell alignment and movement, suggesting a mechanochemical coupling between morphogen gradients and tissue mechanics (Fisher et al., 2017).

The formulation of production/dissipation balance of surface orientational energy density by including informational flux is inspired by Bastock and Strutt (2007), Collu and Mlodzik (2015), and Aw and Devenport (2017). The balance can be expressed as:

$$\frac{dE_{stor}(r,\tau)}{d\tau} = \vec{\boldsymbol{J}}_{\boldsymbol{info}} \cdot \vec{\boldsymbol{\nabla}}\cos\theta + \frac{dE_{diss}(r,\tau)}{d\tau} \tag{3}$$

where $E_{stor}(r,\tau) = E_{stor}^{h} + E_{stor}^{g}$, and $E_{diss}(r,\tau) = E_{diss}^{h} + E_{diss}^{g}$, while $\vec{\boldsymbol{J}}_{\boldsymbol{info}} \cdot \vec{\boldsymbol{\nabla}}\cos\theta$ is the energy source term, $\vec{\boldsymbol{J}}_{\boldsymbol{info}}$ is the informational flux, which measures directed transfer of orientational information in the across of the monolayer. The informational flux can be expressed as:

$$\vec{\boldsymbol{J}}_{\boldsymbol{info}} = -\varphi_{\theta}\vec{\boldsymbol{\nabla}}\cos\theta + \xi\vec{\boldsymbol{\nabla}}a \tag{4}$$

where $\varphi_{\theta}$ is a measure of the transmitting orientational energy per unit length, $\vec{\boldsymbol{\nabla}}\cos\theta$ measures how strongly neighboring cells are misaligned, $\xi$ is the orientational chemomechanical coupling coefficient, which represents the strength of signal-to-orientation transduction, and $a \in [0,1]$ is the dimensionless activity of the signaling pathway. The numerical discretization of the gradients from eq. 4 is shown in **Appendix B**. The first term on the rhs of eq. 4 represents the mechanical orientational resistance generated by cell–cell interactions, while the second term is the active biochemical biasing of orientation. Change in dimensionless activity can be expressed as:

$$\frac{da(r,\tau)}{d\tau} = D_a(\theta)\nabla^2 a - k_d a + S\left(\cos\theta, \vec{\boldsymbol{\nabla}}\cos\theta\right) \tag{5}$$

where $D_a(\theta)$ is the diffusion coefficient, i.e., a “spatial spread” of the signalling activity, $k_d$ is the specific deactivation rate, and $S\left(\cos\theta, \vec{\boldsymbol{\nabla}}\cos\theta\right)$ is the mechanically dependent source term. The local production of activity, expressed by term $S\left(\cos\theta, \vec{\boldsymbol{\nabla}}\cos\theta\right)$, depends on the $\cos\theta$ or orientational gradient $\vec{\boldsymbol{\nabla}}\cos\theta$. The source term is higher for head-on interactions than for glancing interactions. It is in accordance with the fact that the gradient $\vec{\boldsymbol{\nabla}}\cos\theta$ is higher for head-on interactions. Although isotropic diffusion is assumed for the baseline spread of signaling activity, anisotropy emerges through coupling to orientation-dependent source terms and mechanically-induced gradients, resulting in an effectively anisotropic signaling dynamics. For deeper understanding of cell-cell orientational interactions, it is necessary to relate the effective angular modulus to the second virial coefficient. The signalling pathway triggered by cell orientational interactions influences viscoelasticity by influencing memory effects. The heterogeneity resulting from cellular signalling has been investigated in two specific contexts: (1) cells within a uniform population responding to different signals and/or (2) diverse behaviours that may emerge in reaction to the same signals (Blanchard et al, 2019; Petrungaro et al., 2019).

### 3.4 The relationship between the effective angular modulus and 2D second virial coefficient

Because migrating epithelial monolayers are highly dissipative active systems, our formulation utilizes an effective thermodynamic framework mapped onto a non-equilibrium steady state (NESS). Unlike a passive equilibrium state where microscopic currents vanish, this NESS is actively and continuously sustained by internal cell-cell interactions. The lifespan of this local steady state $\tau_{steady}$ corresponds directly to the macroscopic timescale of collective cell migration $\tau_{CCM}$ and is rigorously bounded by a separation of timescales: $\tau_{dc} \leq f_c^{-1} < \tau_{pers} \ll \tau_{steady} < \tau_{CCM}$, where $\tau_{dc}$ is the duration of cell-cell collision (a few minutes), $f_c^{-1}$ is the time between successive collisions (also a few minutes), $\tau_{pers}$ is the persistence time of collective cell migration (a few tens of minutes), $\tau_{steady}$ is the steady-state life spin (a few hours), $\tau_{CCM}$ is the time of collective cell migration (tens of hours). Because $\tau_{dc} \ll \tau_{steady}$ the fast dissipative collision dynamics can be time-averaged and integrated into effective pair-potentials for head-on and glancing interactions $U_{Th}(r,\theta)$, $U_{Tg}(r,\theta)$, mathematically justifying the use of the second virial coefficient in this active system. Consequently, coarse-grained cell-cell interactions within multicellular domains which are in a non-equilibrium steady state can be described by the second virial coefficient.

An effective angular modulus can be formulated from the second virial coefficient. In two dimensions, the effective second virial coefficient represents the excluded surface area $A_{ex}$ generated by cell–cell interactions, $B_{2eff} \sim A_{ex}$. For the cell packing density $n_e > n_{conf}$, orientational interactions perturb cell alignment and cells undergo isotropic migration (Roycraft and Mayor, 2016; Pajic-Lijakovic et al., 2026). In this case, the effective second virial coefficient $B_{2eff}$ is equal to:

$$B_{2eff} = Y_g B_2^g + \left(1 - Y_g\right) B_2^h \tag{6}$$

where $Y_g(n_e, \theta)$ is the fraction of glancing interactions, while $B_2^g$ and $B_2^h$ are the second virial coefficients caused by glancing and head-on interactions respectively expressed in $\mu\text{m}^2$. The second virial coefficients caused by: (i) glancing interactions is $B_2^g = -\frac{1}{2}\int_0^\infty d^2 r_{12} \int_0^{2\pi}\left(e^{-\frac{U_{Tg}(r,\theta)}{\langle e_c\rangle n_e}} - 1\right) d\theta$ and (ii) head-on interactions is $B_2^h = -\frac{1}{2}\int_0^\infty d^2 r_{12} \int_0^{2\pi}\left(e^{-\frac{U_{Th}(r,\theta)}{\langle e_c\rangle n_e}} - 1\right) d\theta$ (where $r_{12}$ is the distance between two cells, $U_{Tg}(r,\theta) = U(r) + U_g(\theta)$ and $U_{Th}(r,\theta) = U(r) + U_h(\theta)$ are the total interaction potential for glancing and head-on interactions, respectively, $U(r)$ is the positional pair-potential, $\langle e_c \rangle n_e \equiv \frac{\langle e_c \rangle}{\langle A_c \rangle}$ is the energy quantum per unit surface, $\langle A_c \rangle$ is the average area per single cell $\langle A_c \rangle = \frac{1}{n_e}$, $n_e$ is the 2D cell packing density, $\langle e_c \rangle$ is the average energy of a single cell $\langle e_c \rangle = \langle e_{el} \rangle + \langle e_a \rangle + \langle e_{contr} \rangle$, $\langle e_{el} \rangle$ is the elastic contribution, $\langle e_a \rangle$ is the contribution of cell-cell adhesion contacts per single cell, $\langle e_{contr} \rangle$ is the contractility contribution, $U_h(\theta)$ is the average storage energy per single head-on collision $U_h(\theta) = \frac{1}{2} k_{eff}^h \Delta\theta^2$, and $U_h(\theta)$ is the average storage energy per single glancing collision $U_g(\theta) = \frac{1}{2} k_{eff}^g \Delta\theta^2$). The second virial coefficients for glancing and head-on interactions satisfy the conditions that $B_2^g \sim \Delta A_{ex}^g$ and $B_2^h \sim \Delta A_{ex}^h$ (where $\Delta A_{ex}^g$ and $\Delta A_{ex}^h$ are the excluded surfaces per single cell caused by glancing and head-on interactions, respectively). Although individual cells may experience multiple simultaneous contacts, the interaction model is formulated in terms of pairwise orientational events that act as coarse-grained building blocks of a dense many-body system. The effect of multiple contacts is incorporated implicitly through density-dependent interaction frequency and effective irreversibility emerging in high-packing regimes. The Lennard-Jones pair-potential, proposed for describing cell-cell positional interactions from Kang et al. (2021), can be expressed as: $U(r) = k\left[\left(\frac{d_{min}}{r}\right)^p - \left(\frac{d_{min}}{r}\right)^q\right]$ (where $k$ is the strength of the interactions per unit surface, $p$ and $q$ are exponents). The proposed values of the exponents are $p = 6$ and $q = 3$ (Kang et al., 2021). The interaction energy can be repulsive, when the local distance $r$ is lower than the minimum distance $d_{min}$. Deforet et al. (2014) proposed that $d_{min} \approx 8\ \mu m$.

### 3.4.1 Interpretation of effective angular moduli for head-on and glancing interactions via the second virial coefficient

Head-on and glancing interaction potential can be expressed as: $U_h(\theta) = \langle e_c \rangle \frac{\Delta A_{ex}^h}{\langle A_c \rangle^2}$ and $U_g(\theta) = \langle e_c \rangle \frac{\Delta A_{ex}^g}{\langle A_c \rangle^2}$, respectively. Consequently, the effective angular modulus for head-on interactions is $k_{eff}^h = 2\langle e_c \rangle n_e^2 \frac{\Delta A_{ex}^h}{\Delta\theta^2}$, while for glancing interactions it is $k_{eff}^g = 2\langle e_c \rangle n_e^2 \frac{\Delta A_{ex}^g}{\Delta\theta^2}$. The ratio between these two moduli is $\frac{k_{eff}^h}{k_{eff}^g} = \frac{\Delta A_{ex}^h}{\Delta A_{ex}^g}$. The changes of excluded area caused by head-on and glancing interactions $\Delta A_{ex}^h$ and $\Delta A_{ex}^g$, respectively can be expressed in the form of Taylor series as:

$$\Delta A_{ex}^h = \frac{1}{2} \frac{d^2 A_{ex}^h}{d\theta^2} /_{\theta=\pi} \Delta\theta^2 \tag{7a}$$

$$\Delta A_{ex}^g = \frac{dA_{ex}^g}{d\theta} \Delta\theta + \frac{1}{2} \frac{d^2 A_{ex}^g}{d\theta^2} /_{\theta=\frac{\pi}{2}} \Delta\theta^2 \tag{7b}$$

Glancing interactions lack left–right symmetry, so the excluded area contains a linear angular term that generates irreversible reorientation and dissipation, while the quadratic term provides transient elastic storage. In contrast to glancing interactions, presented by eq. 7b, head-on interactions are symmetric and the first derivative term $\frac{dA_{ex}^h}{d\theta} \Delta\theta$ tends to zero. The characterization of head-on and glancing interactions based on left-right symmetry is shown in **Figure 5**:

**Figure 5**. Characterization of head-on and glancing interactions based on left-right symmetry. Red arrows represent directions of cell migration, i.e. cell orientations. While cell head-on interactions are symmetric, glancing interactions are asymmetric.

The interactions between cells, particularly in terms of orientation, play a significant role in determining the surface properties and viscoelastic behaviour of migrating epithelial monolayers, a topic that will be explored in greater detail below.

## 4. Impact of orientational cell-cell interactions on epithelial surface tension

Epithelial surface tension represents a free energy per unit multicellular surface, which includes elastic, contractile and cohesion contributions (Koride et al., 2018; Pajic-Lijakovic et al., 2023). The surface tension, in general, is a state variable and can be represented thermodynamically by the Virial equation as function of the packing density of the system constituents (Myers and Freed, 1993). Change in cell packing density directly influences epithelial cohesion. Consequently, the epithelial surface tension can be expressed as (Pajic-Lijakovic et al., 2023):

$$\frac{\gamma_e\,(r,\tau)}{\langle e_c\rangle\, n_e} = 1 - B_{2eff} n_e \qquad (8)$$

where $\gamma_e\,(r,\tau)$ is the epithelial surface tension. The second virial coefficient is higher for a lower cell packing density regime and decreases with epithelial packing density. The product of $B_{2eff}n_e \sim \frac{A_{exc}}{\langle A_c \rangle}$ is: (i) $\frac{A_{ex}}{\langle A_c \rangle} \ll 1$ in the low packing-density regime, (ii) $\frac{A_{ex}}{\langle A_c \rangle}$ increases in the intermediate packing=density regime, and (iii) $\frac{A_{ex}}{\langle A_c \rangle} \to 1$ in the high density regime. Consequently, epithelial surface tension $\gamma_e\,(r,\tau)$: (i) increases with cell packing density almost linearly in the low packing-density regime, (ii) reaches a plateau in the intermediate regime, and (iii) decreases rapidly in the high density regime. Epithelial surface tension varies from a few $\frac{\mathrm{mN}}{\mathrm{m}}$ to a few tens of $\frac{\mathrm{mN}}{\mathrm{m}}$ (Mombach et al., 2005; Stirbat et al., 2013; Guevorkian et al., 2021; Nagle et al., 2022).

Change in epithelial surface tension caused by change in cell packing density induces energy storage and dissipation, which depends on change in the monolayer surface area $\frac{\Delta A_m}{A_m}$ and the rate of its change $\frac{d}{d\tau}\frac{\Delta A_m}{A_m}$, i.e. $\Delta\gamma_e = \Delta\gamma_e\left(\frac{\Delta A_m}{A_m}, \frac{d}{d\tau}\frac{\Delta A_m}{A_m}\right)$. As surface tension increases, particularly in the low-density regime, the capacity for energy storage surpasses that of energy dissipation. Conversely, when surface tension remains approximately constant, as observed in the intermediate regime, energy storage and energy dissipation are nearly equivalent. In contrast, when surface tension decreases, characteristic of the high-density regime, energy storage falls short of energy dissipation. When the distribution of epithelial surface tension and its time change during collective cell migration will be measured, we will be able to postulate constitutive models for dilatational viscoelasticity in the different regimes of cell-packing density.

Epithelial surface tension accompanied by the matrix surface tension contributes to the generation of epithelial-matrix interfacial tension $\gamma_{em}(r,\tau)$ expressed as: $\gamma_{em}(r,\tau) = \gamma_e + \gamma_m - e_a$ (where $\gamma_m$ is the matrix surface tension and $e_a$ is the epithelial-matrix adhesion energy per unit surface) (Pajic-Lijakovic et al., 2024b). Interfacial tension $\gamma_{em}$ further contributes to generation of the isotropic part of the cell compressive stress $\Delta p_c$ expressed by the Young-Laplace equation as: $\Delta p_c = -\gamma_{em}\left(\vec{\boldsymbol{\nabla}} \cdot \vec{\boldsymbol{n}}\right)$ (Marmottant et al., 2009; Pajic-Lijakovic et al., 2024a) (where $\vec{\boldsymbol{n}}$ is a unit normal vector on the epithelial-matrix biointerface) and consequently, stimulates cell head-on interactions. The inhomogeneous distribution of cell packing density during collective cell migration generates a gradient of epithelial surface tension $\vec{\boldsymbol{\nabla}}\gamma_e$ across a multicellular domain. In turn, this gradient $\vec{\boldsymbol{\nabla}}\gamma_e$ generates a gradient of interfacial tension $\vec{\boldsymbol{\nabla}}\gamma_{em}$, even when the matrix surface tension $\gamma_m$ and adhesion energy $e_a$ are homogeneously distributed within part of an epithelial monolayer. The gradient of interfacial tension triggers cell movement from the regions of lower interfacial tension to that of higher interfacial tension by contributing to the generation of shear stress, which has a feedback impact on glancing interactions (Gsell et al., 2023). The phenomenon is known as the Marangoni effect (Karbalaei et al., 2016; Pajic-Lijakovic et al., 2024a). Cell-cell orientational interactions also influence the viscoelasticity of epithelial monolayers caused by collective cell migration.

**5. Impact of irreversible cell-cell interactions on the viscoelasticity of epithelial monolayers**

The viscoelasticity of migrating epithelial monolayers depends on interplay between: cell packing density, mechanisms of cell migration, and the degree of anisotropy. Anisotropic cell migration can be induced by spatially modulated mechanical feedback. Gustafson et al. (2022) demonstrates that this feedback acts as a critical regulatory input to establish long-range gradients of cytoskeletal configurations, such as myosin, which in turn drive global tissue flow patterns. An increase in cell packing density leads to (Pajic-Lijakovic et al., 2026):

- a change in the migration mechanism from: (i) convective - characteristic of the low-density regime, and (ii) diffusion – dominant in the intermediate regime, to (iii) sub-diffusion - characteristic of the high-density regime; and
- a decrease in the degree of anisotropy caused by cell-cell orientational interactions.

The viscoelasticity of epithelial monolayers caused by collective cell migration occurs on two timescales (Marmottant et al., 2009). A timescale of minutes corresponds to the stress relaxation time, while strain change and cell residual stress accumulation occur on a timescale of hours (**Glossary**) (Marmottant et al., 2009; Pajic-Lijakovic and Milivojevic, 2025).

Cell-cell orientational interactions can perturb aligned, anisotropic cell migration depending on the frequency of collisions. Consequently, the degree of anisotropy $\alpha$, which is in the range of $\alpha \in [0,1]$, depends on the frequency $f_c$, such that for $\alpha = 0$, cells undergo isotropic migration, while for $\alpha = 1$, cells perform aligned, anisotropic migration. Consequently, the degree of anisotropy can be expressed as:

$$\alpha = \frac{1}{1+f_c \tau_a} \tag{9}$$

where $\tau_a(n_e)$ is the cell alignment time after cell-cell interaction, which depends on epithelial packing density and satisfies the conditions that: (i) $\tau_a \leq \tau_{RP}$ and (ii) $\tau_a \sim n_e^{-1}$. As a result, the viscoelastic properties of migrating epithelial collectives vary according to the established density regime sizes. Upon further examination, it is essential to delineate the cell strain induced by collective cell migration and its subsequent long-term alterations, followed by the stress generated through the formulation of the constitutive model appropriate to the packing-density regime.

### 5.1 Cell strain caused by collective cell migration

The velocity within multicellular domain can be expressed as: $\vec{\boldsymbol{v}}(r,\tau) = \frac{d\vec{\boldsymbol{u}}_e}{d\tau}$ (where $\vec{\boldsymbol{u}}_e$ is the displacement field caused by collective cell migration). Cell strain depends on the epithelial displacement field and can be expressed as: $\tilde{\boldsymbol{\varepsilon}}(r,\tau) = \frac{1}{2}\left(\vec{\boldsymbol{\nabla}}\vec{\boldsymbol{u}}_e + \vec{\boldsymbol{\nabla}}\vec{\boldsymbol{u}}_e{}^{\boldsymbol{T}}\right)$**.** Cell strain generates stress. The stress-strain constitutive model (**Glossary**) depends on the mechanism of cell migration controlled by the cell-packing density. Upon further examination, it is essential to articulate the contribution of mechanical stress resulting from collective cell migration, as it has a direct impact on the generation of cell strain.

### 5.2 Viscoelasticity in the low-density regime

For the low-density regime, the frequency of collisions $f_c \rightarrow 0$ and the degree of anisotropy $\alpha \rightarrow 1$. It means that cell-cell orientational interactions do not influence the viscoelasticity in this regime. The viscoelastic properties of epithelial monolayers, resulting from collective cell migration for $n_e \leq n_{conf}$, fulfill the following criteria: (i) cell stress is capable of relaxation when subjected to constant strain (Khalilgharibi et al., 2019), (ii) the relaxation of stress follows an exponential pattern (Marmottant et al., 2009), corresponding to the behavior of a linear viscoelastic solid, (iii) stress relaxation occurs over minutes (Khalilgharibi et al., 2019), and (iv) the residual stress in cells (**Glossary**) is associated with the corresponding strain and changes over hours (Serra-Picamal et al., 2012; Notbohm et al., 2016). The capacity for strain to relax exponentially under conditions of constant (or zero) stress (Marmottant et al., 2009) suggests a specific linear constitutive model, namely, the Zener model applicable to viscoelastic solids (Pajic-Lijakovic et al., 2026). The strain relaxes over hours under constant stress conditions (Marmottant et al., 2009). Stress relaxes within successive short-time stress relaxation cycles, while strain caused by collective cell migration changes slowly, over hours. Consequently, stress changes from its initial value $\widetilde{\boldsymbol{\sigma}}^{\mathbf{0}}{}_i(r,t,\tau)$ towards the cell residual stress $\widetilde{\boldsymbol{\sigma}}_{Ri}(r,\tau)$ per constant strain $\tilde{\boldsymbol{\varepsilon}}_{\boldsymbol{i}}(r,\tau)$ within a single stress relaxation cycle (where $i \equiv S, N$, $S$ is the shear component, $\sigma_{xy}$ and $N$ are the normal components $\sigma_{xx}$ and $\sigma_{yy}$, $t$ is the short timescale of minutes, and $\tau$ is the long timescale of hours)**.** Then strain increases steeply from $\tilde{\boldsymbol{\varepsilon}}_{\boldsymbol{i}}(r,\tau)$ to $\tilde{\boldsymbol{\varepsilon}}_{\boldsymbol{i}} + \Delta\tilde{\boldsymbol{\varepsilon}}_{\boldsymbol{i}}$ and the stress relaxes to this value within the next stress relaxation cycle.

The level of generated stress is higher in the context of anisotropic cell migration relative to isotropic migration, assuming identical cell-packing density. It is consistent with the understanding that all strain components $\varepsilon_{xy}$, $\varepsilon_{xx}$, and $\varepsilon_{yy}$ simultaneously contribute to the generation of each stress component in the context of anisotropic cell migration (Pajic-Lijakovic et al., 2026). In this case, despite epithelial monolayers show[ng linear viscoelastic behaviour, a simple shear motion of epithelial monolayers generates the first normal stress difference $N_1 = \sigma_{xx} - \sigma_{yy}$. However, when the cell migration is isotropic: (i) the shear strain component $\varepsilon_{xy}$ contributes only to the shear stress $\sigma_{xy}$, (ii) the normal strain component $\varepsilon_{xx}$ contributes to the normal stress component $\sigma_{xx}$, and (iii) the normal strain component $\varepsilon_{yy}$ contributes to the normal stress component $\sigma_{yy}$. In this line, anisotropic cell migration, which corresponds to: a cell packing density range of: $n_e \leq n_{conf}$ and a cell velocity range of: $0.1 < \vec{\boldsymbol{v}}_{\boldsymbol{c}} < \sim 1 \frac{\mu\text{m}}{\text{min}}$, can be described by the modified Zener model when stress and strain are measured. To account for stress generation associated with anisotropic collective migration, an additional constitutive term was introduced by Pajic-Lijakovic et al. (2026):

$$\widetilde{\boldsymbol{\sigma}}_i + \tau_{Rs}\dot{\widetilde{\boldsymbol{\sigma}}}_i = \widetilde{\boldsymbol{\sigma}}^{\mathbf{0}}{}_i\left(\tilde{\boldsymbol{\varepsilon}}_i, \dot{\tilde{\boldsymbol{\varepsilon}}}_i\right) + \alpha\Delta\widetilde{\boldsymbol{\sigma}}_i\left(\tilde{\boldsymbol{\varepsilon}}_i, \dot{\tilde{\boldsymbol{\varepsilon}}}_i\right) \quad (10)$$

where $i \equiv S, N$, $S$ is the shear component of stress $\sigma_{xy}$ and strain $\varepsilon_{xy}$, while $N$ are the normal components of stress $\sigma_{xx}$, $\sigma_{yy}$ and strain $\varepsilon_{xx}$ and $\varepsilon_{yy}$, $\dot{\widetilde{\boldsymbol{\sigma}}}$ is the rate of stress change $\dot{\widetilde{\boldsymbol{\sigma}}}_i = \frac{d}{dt}\widetilde{\boldsymbol{\sigma}}_i$, $\widetilde{\boldsymbol{\sigma}}^{\mathbf{0}}{}_i$ is the isotropic stress $\widetilde{\boldsymbol{\sigma}}^{\mathbf{0}}{}_i\left(\tilde{\boldsymbol{\varepsilon}}_i, \dot{\tilde{\boldsymbol{\varepsilon}}}_i\right) = E_{\boldsymbol{i}}\tilde{\boldsymbol{\varepsilon}}_{\boldsymbol{i}} + \eta_i\dot{\tilde{\boldsymbol{\varepsilon}}}_{\boldsymbol{i}}$**,** $E_{\boldsymbol{i}} \equiv E_{\boldsymbol{s}}$ is the shear elastic modulus, $E_{\boldsymbol{i}} \equiv E_{\boldsymbol{N}}$ is the Young's modulus, $\eta_i \equiv \eta_s$ is the shear viscosity, and $\eta_i \equiv \eta_N$ is the normal viscosity, $\tau_{Rs}$ is the stress relaxation time, which corresponds to a timescale of minutes, and $\Delta\tilde{\boldsymbol{\sigma}}_i$ is the anisotropic contribution to stress $\Delta\tilde{\boldsymbol{\sigma}}_i(\tilde{\boldsymbol{\varepsilon}}_i, \dot{\tilde{\boldsymbol{\varepsilon}}}_i) = \tilde{\boldsymbol{\sigma}}_i - \frac{1}{d} tr(\tilde{\boldsymbol{\sigma}}_i)\tilde{\boldsymbol{I}}$, $d \equiv 2$ is the dimensionality of space, $t$ is the timescale of minutes, $\tau$ is the timescale of hours, and $\tilde{\boldsymbol{I}}$ is the unit tensor. The Young's modulus of contractile epithelial MDCK monolayers is approximately ~33 kPa (Schulze et al., 2017). The maximum compressive stress generated during collective migration of confluent epithelial MDCK monolayers was ~300 Pa (Notbohm et al., 2016), while shear stress was ~100 Pa (Tambe et al., 2013). An additional increase in cell packing density leads to an increase in the magnitudes of both the compressive and shear stress components.

Energy dissipation in this regime occurs over minutes during stress relaxation toward the residual stress, which is elastic for the Zener model. The main factor contributing to energy dissipation in this context is the remodeling of cell-cell and cell-matrix adhesion, which facilitates collective cell migration, as opposed to orientational interactions between cells (Pajic-Lijakovic and Milivojevic, 2025). The energy dissipation per single stress relaxation cycle within the time increment $\Delta\tau$ can be expressed as: $\Delta E_{diss}^T = \int(\tilde{\boldsymbol{\sigma}}^{\boldsymbol{0}}{}_i - \tilde{\boldsymbol{\sigma}}_{Ri}) : \dot{\tilde{\boldsymbol{\varepsilon}}}_i \, \Delta\tau dV$, while the energy storage is: $\Delta E_{stor}^T = \int \tilde{\boldsymbol{\sigma}}_{Ri} : \dot{\tilde{\boldsymbol{\varepsilon}}}_i \, \Delta\tau dV$ (where $\Delta V$ is the volume increment of epithelial monolayer: $\Delta V = \Delta A_m l_c$, and $l_c$ is the average height of the epithelial monolayer) (Pajic-Lijakovic and Milivojevic, 2025).

Consequently, the fractions of total energy storage $X_{stor}^{h+g}$ and dissipation $X_{diss}^{h+g}$ caused by head-on and glancing interactions during the time increment $\Delta\tau$ in this regime satisfy the conditions: (i) $X_{stor}^{h+g} \to 0$ and (ii) $X_{diss}^{h+g} \to 0$ (where the fraction of total energy storage caused by head-on and glancing interactions during the time increment $\Delta\tau$ is $X_{stor}^{h+g} = \frac{(\Delta E_{stor}^h + \Delta E_{stor}^g)\frac{1}{l_c}}{\Delta E_{stor}^T}$, $l_c$ is the average thickness of the epithelial monolayer, and the fraction of total energy dissipation is $X_{diss}^{h+g} = \frac{(\Delta E_{diss}^h + \Delta E_{diss}^g)\frac{1}{l_c}}{\Delta E_{diss}^T}$). The energy storage and dissipation caused by cell head-on and glancing interactions were formulated in the **Section 3.1.** An increase in cell packing density changes the mechanisms of cell migration and viscoelasticity of epithelial monolayers.

**5.3 Viscoelasticity in the intermediate-density regime**

In the intermediate-density regime, the frequency $f_c$ and fraction of irreversible NE-11 interactions increases, leading to disturbances in cell alignment. The frequency of orientational interactions satisfies the condition that $f_c < \frac{1}{\tau_{RP}}$. Consequently, cells have enough time to repolarise after head-on interactions. It means that after interactions, the majority of cells continue their migration, while irreversible interactions such as: NE-10, NE-01, NE-00, and NE-X can be neglected. Thus, viscoelasticity is linked to isotropic cell migration, while the migration mechanism is characterized as linear and diffusion-based. The stress relaxation time exceeds the average cell-cell collision times. For the conditions that: $\tau_{Rs} \gg f_c^{-1}$ and $\alpha \sim 0$, the constitutive model presented by eq. 10 reduces to the Kelvin-Voigt model, which corresponds to the cell packing density range of $n_{conf} < n_e < n_j$ and cell velocity range of:

$\vec{\boldsymbol{v}}_{\boldsymbol{c}} \sim 10^{-3} - 10^{-2} \frac{\mu m}{min}$ . In this limit, the anisotropic contribution becomes negligible and the stress relaxation term does not affect the constitutive response at the interaction timescale, leading to the standard Kelvin–Voigt form. The model is expressed as (Pajic-Lijakovic et al., 2026):

$$\tilde{\boldsymbol{\sigma}}_i = E_{\boldsymbol{i}} \tilde{\boldsymbol{\varepsilon}}_{\boldsymbol{i}} + \eta_{\boldsymbol{i}} \dot{\tilde{\boldsymbol{\varepsilon}}}_{\boldsymbol{i}} \tag{11}$$

where the elastic contribution to stress $\tilde{\boldsymbol{\sigma}}_{el\,i}$ is equal to $\tilde{\boldsymbol{\sigma}}_{el\,i} = E_{\boldsymbol{i}} \tilde{\boldsymbol{\varepsilon}}_{\boldsymbol{i}}$ and the viscous contribution to stress $\tilde{\boldsymbol{\sigma}}_{vis\,i}$ is $\tilde{\boldsymbol{\sigma}}_{vis\,i} = \eta_{\boldsymbol{i}} \dot{\tilde{\boldsymbol{\varepsilon}}}_{\boldsymbol{i}}$. In this case, energy dissipation occurs on a time scale of hours as a consequence of cell-cell orientational interactions (Pajic-Lijakovic and Milivojevic, 2025). The energy storage can be expressed as: $\Delta E_{stor}^{T} = \int \tilde{\boldsymbol{\sigma}}_{el\,i} : \dot{\tilde{\boldsymbol{\varepsilon}}}_{\boldsymbol{i}} \, \Delta\tau dV$, while the energy dissipation can be written: $\Delta E_{diss}^{T} = \int \tilde{\boldsymbol{\sigma}}_{vis\,i} : \dot{\tilde{\boldsymbol{\varepsilon}}}_{\boldsymbol{i}} \, \Delta\tau dV$. The fraction of total energy storage and dissipation caused by head-on and glancing interactions in this regime satisfies the conditions that: (i) $X_{stor}^{h+g} < 1$, (ii) $X_{diss}^{h+g} < 1$, and (iii) $X_{stor}^{h+g} > X_{diss}^{h+g}$. Energy storage and dissipation come from both intrinsic cell mechanics such as cortical elasticity, contractility, and cell-cell adhesion, as well as from orientational interactions. Head-on and glancing interactions modulate viscoelasticity, but do not dominate in this regime.

### 5.4 Viscoelasticity in the high-density regime

In the high-density regime near cell jamming, the frequency of interactions between cells is $f_c > \frac{1}{\tau_{RP}}$. It means that cells do not have enough time to repolarise after head-on interactions, which favours irreversible NE-10, NE-01, NE-00 interactions responsible for the damping of cell migration (Garcia et al., 2015; Pajic-Lijakovic et al., 2024a). Shear stress under high cell packing density can lead to live cell extrusion during glancing interactions, described as NE-X (Saw et al., 2017; Pajic-Lijakovic et al., 2024a). After interactions, the majority of cells stop their migration. Cells undergo damped isotropic migration, characterized by a non-linear sub-diffusion mechanism that incorporates fractional derivatives in its formulation. Under these conditions, the cellular system stores and dissipates energy anomalously. Consequently, the first term on the right-hand side of Eq. 11, which represents regular elastic energy storage, is negligible, while the second term on the right-hand side of Eq. 11 describes anomalous energy storage and dissipation. The proposed non-linear constitutive model, which satisfies these conditions and corresponds to the cell packing density $n_e \to n_j$ and cell velocity $\vec{\boldsymbol{v}}_{\boldsymbol{c}} \to 0$, is expressed as (Pajic-Lijakovic et al., 2025):

$$\tilde{\boldsymbol{\sigma}}_{\boldsymbol{i}} = \eta_{\beta} D^{\beta}(\tilde{\boldsymbol{\varepsilon}}_{\boldsymbol{i}}) \tag{12}$$

where $\eta_{\beta}$ is the effective modulus, $D^{\beta}(\boldsymbol{\varepsilon}_{\boldsymbol{i}})$ is the fractional derivative, and $\beta$ is the order of fractional derivatives, which is equal to $\beta = 1 - \sum_k P_k$, while $k \equiv NE-10, NE-01, NE-00, NE-X$. When $\sum_k P_k \to 1$, than $\beta \to 0$, which is characteristic of the cell jamming state. Caputo's definition of the fractional derivative of a function $(\tilde{\boldsymbol{\varepsilon}}_{\boldsymbol{i}})$ is used and expressed as: $D^{\beta} \varepsilon_i = \frac{1}{\Gamma(1-\beta)} \frac{d}{d\tau} \int_0^{\tau} \frac{\varepsilon_i(r,\tau')}{(\tau-\tau')^{\beta}} d\tau'$, and $\Gamma(1-\beta)$ is a gamma function (Podlubny, 1999). The L1-approximation scheme for computing the Caputo derivative is presented in **Appendix C**.

In the high-density regime near jamming, irreversible head-on (NE 10, 01, 00) and glancing (X) interactions dominate both energy storage and dissipation, effectively controlling the viscoelastic response of the monolayer. Consequently, the fractions of total energy storage $X_{stor}^{h+g}$ and dissipation $X_{diss}^{h+g}$ caused by head-on and glancing interactions during the time increment $\Delta\tau$ in this regime satisfy the conditions: (i) $X_{stor}^{h+g} \to 1$ and (ii) $X_{diss}^{h+g} \to 1$. Given the fact that physical parameters, including cell packing density, velocity, strain, epithelial surface tension, and cell stress, are all interconnected and influenced by orientational interactions in the higher density regime where $n_e > n_{cong}$, it is essential to explore this relationship in greater detail.

## 6. Interrelationship between model parameters

Orientational energy stored and dissipated during head-on and glancing interactions does not constitute an independent energetic contribution. Rather, it represents changes in the elastic, contractile, and cohesive components entering the macroscopic viscoelastic response of the migrating epithelial monolayer and its surface properties. The interrelationships between the physical parameters relating cell head-on and glancing interactions with viscoelasticity and the surface properties of epithelial monolayers are shown schematically in **Figure 6**:

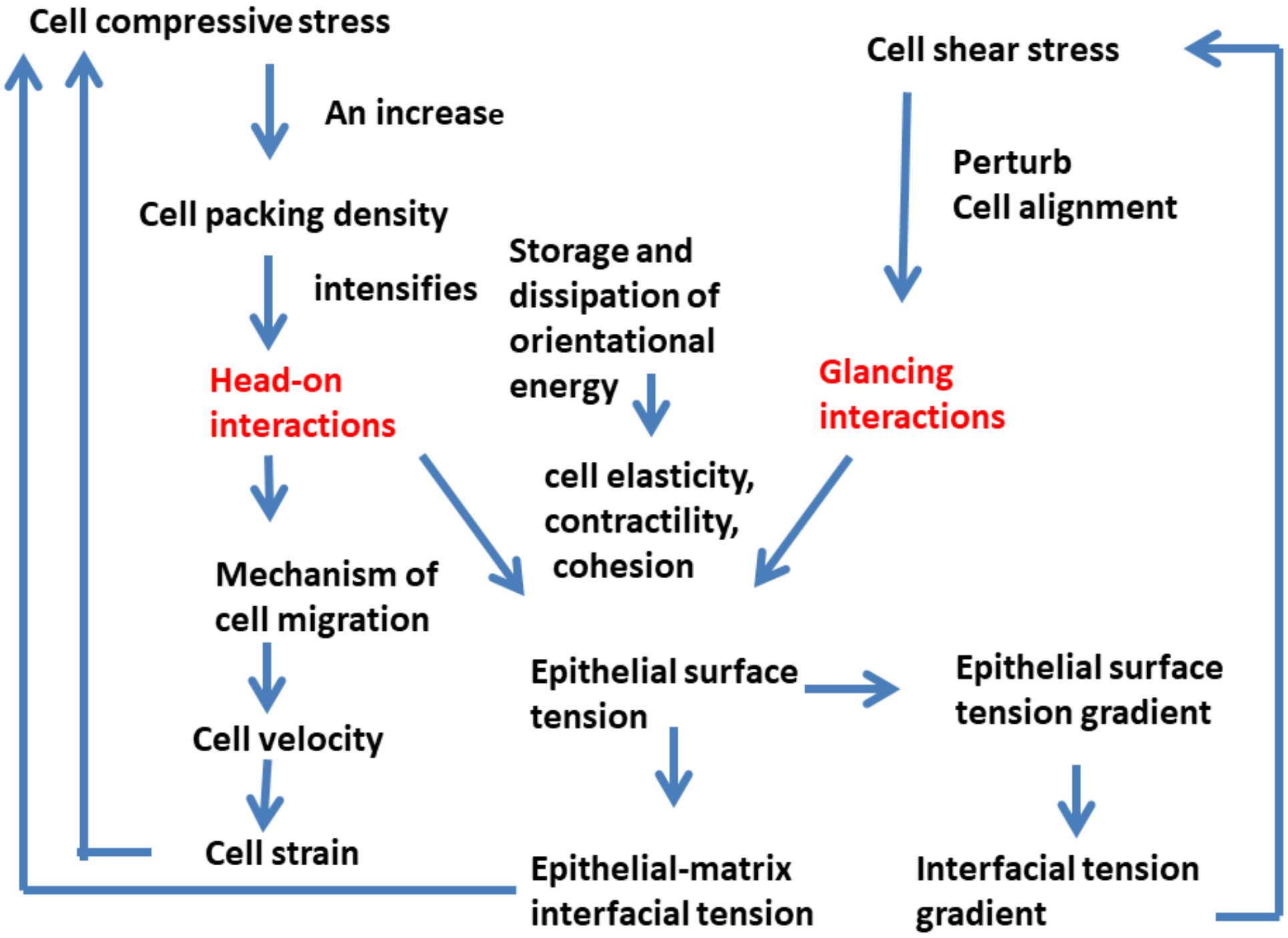

**Figure 6**. Interrelationships between physical parameters relating cell head-on and glancing interactions with viscoelasticity and surface properties of epithelial monolayers such as: cell velocity, strain,

mechanical stress, cell packing density, epithelial surface tension, epithelial-matrix interfacial tension, and interfacial tension gradient.

No explicit quantum or electronic dynamics are included in the present model; their effects are implicitly incorporated through effective coarse-grained parameters. The model parameters, measurement modalities, and numerical estimators for epithelial monolayer dynamics are summarized in **Table 2**:

**Table 2**. Summary of model parameters, measurement modalities, and numerical estimators for epithelial monolayer dynamics

| Parameter | Symbol | Unit | Range | Modality | Estimator |
|---|---|---|---|---|---|
| Probability | $P_i$ | - | $[0,1]$ | Live imaging | multinomial or hidden-Markov model |
| Collision frequency | $f_c$ | $s^{-1}$ | From tens of minutes to hours | Trajectory analysis | Pairwise statistics |
| Repolarisation time | $\tau_{RP}$ | s | From tens of minutes to hours | Polarity markers | Autocorrelation |
| Effective angular moduli | $k_{eff}$ | $\frac{N}{m}$ | $10^{-3} - 1 \frac{mN}{m}$ | AFM indentation/traction force microscopy | iFEM Calibration |
| Excluded area | $\Delta A_{ex}^{g}$<br>$\Delta A_{ex}^{h}$ | $m^2$ | $10^{-3} - 10^{-2}\ \mu m^2$<br>$10^{-2} - 10^{-1}\ \mu m^2$ | Segmented geometry | Pair-correlation |
| Morphogen Diffusivity | $D_a$ | $\frac{m^2}{s}$ | $\sim 10^{-12}\ \frac{m^2}{s}$ | Fluorescent reporters | Inverse R-D problem |
| Epithelial surface tension | $\gamma_e$ | $\frac{N}{m}$ | From a few $\frac{N}{m}$ to a several tens of $\frac{N}{m}$ | Micropipette aspiration/laser ablation | Yang-Laplace fit |
| Mechanical stress | $\tilde{\boldsymbol{\sigma}}_S$ and $\tilde{\boldsymbol{\sigma}}_N$ | Pa | From a few tens of Pa to several hundreds of Pa | Traction force microscopy/Monolayer stress microscopy | Finite element integration |

where $i$ represents various reversible and irreversible head-on and glancing interactions

The parameters listed in **Table 2** include stochastic, geometric, and constitutive quantities operating at different levels of coarse-graining. They control interaction statistics (probabilities, collision frequency), mechanical responses (angular moduli, surface tension, stress), and transport or structural constraints (diffusivity, excluded area). These parameters define the input structure of the model and are used to generate quantitative predictions of tissue dynamics. The model is evaluated within an out-of-sample predictive framework, in which parameters are inferred from early-time dynamics and used to predict independent later-time behaviour. Observable phenomena that can be directly tested encompass density-dependent transition probabilities, velocity fields, correlation properties, dynamics of stress relaxation, trends in surface tension, the onset of jamming, and transitions in migration anisotropy. Model performance can be assessed through predictive error in unseen time windows and by comparison across alternative constitutive formulations, including Zener, Kelvin–Voigt, and fractional viscoelastic models, with uncertainty propagation included in all predicted quantities.

Beyond point estimation, the field is increasingly adopting rigorous frameworks to propagate uncertainty from experimental inputs—such as cell segmentation and traction stress reconstruction—into tissue-level predictions. Modern workflows utilize Bayesian inverse uncertainty quantification (UQ) to define parameter credible intervals rather than single fixed values, often using Gaussian processes or auto-encoders as probabilistic surrogate models to manage the high computational cost of these simulations (Deshpande et al., 2025). Furthermore, statistically enhanced learning (SEL) offers a promising route for model selection and predictive validation by boosting the performance of learning algorithms through the integration of intrinsic statistical features (Felice et al., 2025). Together, these methodologies represent a necessary transition toward more robust and reproducible validation of collective cell migration models.

Cell compressive and shear stress components, generated during collective cell migration, are responsive for cell-cell orientational interactions. Cell compressive stress causes an increase in cell packing density that stimulates cell head-on interactions. Conversely, an increase in shear stress contributes to the formation of topological defects and induces swirling motion in multicellular systems, thereby enhancing glancing interactions between them (Saw et al., 2017; Lin et al., 2018). An increase in cell packing density, caused by an increase in compressive stress, has a feedback impact on compressive stress itself and the rate of its change by influencing the mechanism of cell migration and consequently, cell velocity and corresponding strain. Stress-strain constitutive models describe the mechanisms of energy storage and dissipation. An increase of cell packing density influences epithelial surface tension. Epithelial surface tension accompanied by the surface tension of substrate matrix and cell-matrix adhesion energy contributes to epithelial-matrix interfacial tension. Interfacial tension in turn contributes to generation of an isotropic part of compressive stress based on the Young-Laplace equation, while the deviatoric contribution is generated by collective cell migration (Pajic-Lijakovic et al., 2024a,b). Both contributions intensify cell head-on interactions. Gradient of epithelial surface tension, caused by inhomogeneous distribution of cell packing density, leads to generation of epithelial-matrix interfacial tension gradient and on that bases contribute to an increase in the cell shear stress, which stimulates cell glancing interactions (Pajic-Lijakovic et al., 2024b). The phenomenon that interfacial tension gradient triggers cell migration represents a part of the Marangoni effect (Gsell et al., 2023; Pajic-Lijakovic et al., 2024a).

Cell–cell orientational interactions may perturb collective alignment and promote a transition from anisotropic to isotropic cell migration, particularly at high cell packing densities quantified by the degree of anisotropy (Pajic-Lijakovic et al., 2026). These interactions induce cell signalling, which has a feedback impact on elasticity, contractility, and cohesion of migrating epithelial monolayers and influence the mechanism of cell migration. At higher cell packing density, two outcomes are possible: cell jamming and live cell extrusion (**Glossary**). Head-on and glancing interactions were characterized by effective angular modulus, energy storage and dissipation, frequency of collisions, and excluded surface per single cell. Physical aspects of these complex phenomena were discussed in terms of the set of interrelated physical parameters such as: cell packing density, epithelial surface tension, degree of anisotropy, and mechanical stress.

Future extensions of this framework may address disease-related alterations in cellular mechanics arising from mitochondrial dysfunction and neurodegeneration, where metabolic constraints modify effective mechanical parameters (Grünewald et al., 2007; Gabbert et al., 2022).

**7. Conclusion**

The aim of this work has been to examine how orientational interactions influence key aspects of collective cell migration and contribute to its overall efficiency. In this context, we have introduced a classification of cell–cell interactions into reversible (effectively elastic) and irreversible (dissipative) processes, providing a coarse-grained framework for describing collision-driven reorganization in epithelial monolayers. We propose that reversible interactions are associated with transient storage of orientational potential energy, whereas irreversible head-on and glancing interactions contribute to energy dissipation and changes in cell orientation and alignment. Reversible interactions are associated with transient storage of orientational potential energy, whereas irreversible head-on and glancing interactions contribute to energy dissipation and changes in cell orientation and alignment.

During both types of interaction, the rotation and realignment of cells transiently induce frictional effects within the cytoskeleton and adhesion structures but, for reversible interactions, these effects are reversible, allowing the cells to restore the original collision angle while storing orientational energy. The cumulative effect of these interactions regulates cell elasticity, contractility, and cell–cell adhesion, thereby modifying the surface properties and effective viscoelastic response of the epithelial monolayer. Increasing cell packing density shifts the balance from elastic storage to dissipative remodeling, producing density-dependent changes in migration mechanisms, epithelial surface tension, and viscoelasticity. Within this framework, orientational interactions are not treated as an independent energetic contribution, but rather as a microscopic mechanism linking cell-scale collisions to emergent tissue-scale mechanics.

Specifically, we (i) introduce a description of reversible and irreversible head-on and glancing interactions in terms of an orientational potential, as a means of organizing their contributions to energy storage and dissipation; (ii) propose an effective second virial coefficient linking excluded surface area to cell packing density and orientational interactions within a coarse-grained approximation; and (iii) explore a density-dependent relationship between epithelial surface tension, viscoelastic response, and interaction-induced remodeling.

Finally, we introduced dimensionless ratios that quantify the relative fractions of total energy storage and dissipation attributable to orientational interactions, revealing that these contributions are partial at intermediate densities but become dominant at higher cell packing density near jamming. Together, these results provide a mechanobiological framework linking microscopic interaction dynamics, cell-state remodelling, and macroscopic epithelial mechanics, and suggest a possible route through which cell-scale interactions may contribute to tissue-scale behaviour.

**Declarations**

**Ethics approval and consent to participate**: Not applicable.

**Consent for publication**: All authors have agreed to publish this manuscript.

**Availability of data and material**: Not applicable.

**Conflict of interest**: The authors report there is no conflict of interest.

**Funding**: This work was supported in part by the Engineering and Physical Sciences Research Council, United Kingdom (grant number EP/X004597/1) and by the Ministry of Science, Technological Development and Innovation of the Republic of Serbia (Contract No. 451-03-34/2026-03/ 200135).

**Authors' contributions**: The authors contributed equally to manuscript preparation.

**Appendix A: Coarse-Graining Procedure and Numerical Implementation of the Continuum Fields**

To construct a continuous velocity field from discrete cell positions, we define a coarse-grained representation of the cell velocities as:

$$\vec{v} = \frac{\sum_i \vec{v}_i W(r-r_i)}{\sum_i W(r-r_i)} \qquad \text{(A1)}$$

where $W(r-r_i)$ the coarse-graining (smoothing) kernel is expressed as: $W(r-r_i) = \frac{1}{2\pi l} e^{-\frac{|r-r_i|^2}{2l^2}}$, and $l$ is the average diameter of an individual cell.

**Appendix B**: The discretisation of gradients from eqs. 3 and 4

The numerical discretization of the information flux gradients can be expressed as:

$$\frac{\partial cos\theta}{\partial m} = \frac{cos\theta_{i+1,j} - cos\theta_{i-1,j}}{2\Delta m} \text{ and } \frac{\partial a}{\partial m} = \frac{cosa_{i+1,j} - cosa_{i-1,j}}{2\Delta m} \qquad \text{(B1)}$$

where $m \equiv x, y$.

**Appendix C**: The L1-approximation scheme for computing the Caputo derivative

The L1-scheme is the commonest  numerical method for approximating the Caputo fractional derivative of order 0$< \beta < 1$. The scheme approximates the function $f(t)$ using piecewise linear interpolation over each small time interval $[t_k, t_{k+1}]$. For a uniform time-step $\Delta t$, the L1-approximation of the Caputo derivative at time $t_n$ is defined as:

$$D^{\beta}\big(f(t_n)\big) \approx \frac{(\Delta t)^{-\beta}}{\Gamma(2-\beta)} \sum_{k=0}^{n-1} [f(t_{k+1}) - f(t_k)]\, a_{n-1-k} \qquad \text{(C1)}$$

where $a_j$ is the waiting coefficient equal to: $a_j = (j+1)^{1-\beta} - j^{1-\beta}$. We assume a quiescent history, meaning that for all time $t \leq 0$ the function $f(t)$ is constant. Because fractional derivatives accumulate history, small errors in each time-step can add up over time. We perform a sensitivity analysis by halving the time-step $\frac{\Delta t}{2}$ and ensuring the total 'memory' of the system remains consistent and the solution converges.